\documentclass[prd,twocolumn,superscriptaddress,altaffilletter,showpacs,nofootinbib]{revtex4}
\usepackage[dvips]{graphicx}
\usepackage{amsmath}

\newcommand{\be}{\begin{equation}}
\newcommand{\ee}{\end{equation}}
\newcommand{\bea}{\begin{eqnarray}}
\newcommand{\eea}{\end{eqnarray}}

\begin{document}

\title{No compelling cosmological models come out of magnetic universes which are based in nonlinear electrodynamics}

\author{Ricardo Garc\'ia-Salcedo}\email{rigarcias@ipn.mx}\affiliation{Centro de Investigacion en Ciencia Aplicada y Tecnologia Avanzada - Legaria del IPN, M\'exico D.F., CP 11500, M\'exico.}

\author{Tame Gonzalez}\email{tamegc72@gmail.com}\affiliation{Departamento de Ingenier\'ia Civil, Divisi\'on de Ingenier\'ia, Universidad de Guanajuato, Guanajuato, CP 36000, M\'exico.}

\author{Israel Quiros}\email{iquiros6403@gmail.com}\affiliation{Departamento de Ingenier\'ia Civil, Divisi\'on de Ingenier\'ia, Universidad de Guanajuato, Guanajuato, CP 36000, M\'exico.}\affiliation{Departamento de Matem\'aticas, Centro Universitario de Ciencias Ex\'actas e Ingenier\'ias, Universidad de Guadalajara, Guadalajara, CP 44430, Jalisco, M\'exico.}

\date{\today}

\begin{abstract}
Here we investigate the cosmic dynamics of Friedmann-Robertson-Walker universes -- flat spatial sections -- which are driven by nonlinear electrodynamics (NLED) Lagrangians. We pay special attention to the check of the sign of the square sound speed since, whenever the latter quantity is negative, the corresponding cosmological model is classically unstable against small perturbations of the background energy density. Besides, based on causality arguments, one has to require that the mentioned small perturbations of the background should propagate at most at the local speed of light. We also look for the occurrence of curvature singularities. Our results indicate that several cosmological models which are based in known NLED Lagrangians, either are plagued by curvature singularities of the sudden and/or big rip type, or are violently unstable against small perturbations of the cosmological background -- due to negative sign of the square sound speed -- or both. In addition, causality issues associated with superluminal propagation of the background perturbations may also arise.
\end{abstract}


\maketitle

\section{Introduction}\label{sec-intro}

There are two problems that bother cosmologists more than any other: (i) the initial cosmological -- big bang -- singularity, and (ii) the current accelerated pace of the cosmic expansion. While for the latter issue there are plenty of models which have been more or less successfully tested against the existing observational data \cite{de-models, odintsov}, for the former problem there are only a few sound proposals which are untested due to the lack of data about the primordial state of the universe. Among these models, perhaps the better known are the pre-big bang (PBB) cosmology scenario \cite{pbb} and the ekpyrotic universe \cite{ekpyrotic}.

Homogeneous and isotropic non singular Friedmann-Robertson-Walker (FRW) cosmological models can be obtained also by considering local covariant and gauge-invariant Lagrangian generalizations of Maxwell electrodynamics which have been dubbed as nonlinear electrodynamics (NLED) theories \cite{klippert, novello-prd-2004, novello-cqg-2007, born-infeld, novello-prd-2012}. In a cosmological setting these theories have been explored mainly within the so called ``magnetic universe'' approximation. According to this approximation, in the early universe, at temperatures above and below -- depending on the model of inflation -- the electroweak (EW) scale [$10^{12}-10^{16}$ GeV], where the spacetime is filled with an equilibrium primordial plasma of elementary particles \cite{shaposhnikov-prd-1998, joyce-prl-1997}, we may assume that only the average of the magnetic field squared $B^2$ survives.\footnote{This amounts to neglect the bulk viscosity terms in the electric conductivity of the primordial plasma \cite{klippert}.} It is known that large enough magnetic fields [$\sim 10^{22}$ Gauss at temperatures $T\sim 100$ GeV] coherent on different scales, may be generated in the primordial universe due to several mechanisms, such as:\footnote{The presence of large scales magnetic fields in our observed universe is a well established observational fact \cite{mag-f}.} first-order quark-hadron phase transition \cite{olinto}, EW phase transition \cite{kibble}, primordial lepton asymmetry \cite{joyce-prl-1997}, parametric amplification of quantum vacuum fluctuations of some primordial gauge field \cite{grishchuk}, breakdown of conformal invariance due to coupling of the inflaton to the Maxwell term in the chaotic inflationary model \cite{ratra} (see also \cite{turner}), and coupling of the photon either to the dilaton \cite{veneziano} or to the graviphoton \cite{gasperini'}, among others. In this context the study of a magnetic universe based in NLED theories looks like a plausible possibility to seek for interesting cosmological effects. A prototype of a NLED theory is provided by the Born-Infeld (BI) Lagrangian \cite{born-infeld}: 

\bea {\cal L}=-\gamma^2\left(\sqrt{1+F/2\gamma^2}-1\right),\label{l-b-i}\eea where $\gamma$ is a free parameter and the electromagnetic invariant\footnote{The electromagnetic tensor is defined as $F_{\mu\nu}:=A_{\nu,\mu}-A_{\mu,\nu}$.} 

\bea F\equiv F^{\mu \nu}F_{\mu \nu }=2({\bf B}^2-{\bf E}^2).\label{f-inv}\eea The motivation of the authors was to have regular field configurations without singularities. The gravitational field was not included in their analysis. If one introduces gravitational effects through the theory of general relativity, a drawback of Born-Infeld proposal unfolds: there is no place for a regular cosmological scenario with the combined effects of gravity and NLED.\footnote{See, however, the reference \cite{nora-2000} where it was demonstrated that the Born-Infeld theory is singularity free in Bianchi spacetimes.} 

In the reference \cite{novello-prd-2012}, motivated by the original Born-Infeld's idea of having regular field configurations with the electromagnetic (EM) field bounded -- this time in a magnetic universe -- the authors focused in a modification of the Lagrangian (\ref{l-b-i}) by the inclusion of a term quadratic in the field $F$ within the square root

\bea {\cal L}=-\gamma^2\,W^{1/2},\;W:=1+\frac{F}{2\gamma^2}-\alpha^2 F^2,\label{l-model}\eea where $\alpha$ is another free constant and, as seen, the term $\propto\gamma^2$ in (\ref{l-b-i}) has been removed. This modified theory does not have the standard (classical) linear -- weak field -- Maxwell limit, since at $\alpha\rightarrow 0$, $\gamma\rightarrow\infty$, it includes the effects of the (vacuum) zero-point fluctuations of the EM fields due to the constant term in $${\cal L}\approx-\frac{F}{4}-\gamma^2.$$ An alternative theory which does have the classical Maxwell limit it is given by the Lagrangian density

\bea {\cal L}=-\gamma^2\left(W^{1/2}-1\right),\label{l-model'}\eea with $W$ defined as in Eq. (\ref{l-model}). 

Magnetic universes have been investigated also within the context of NLED theories which are given by the much simpler Lagrangian density \cite{klippert} 

\bea {\cal L}=-\frac{1}{4}F+\alpha F^2,\label{l-f2}\eea where the nonlinear term $\propto F^2$ may cause the universe to bounce thus avoiding the initial (big bang) singularity. Lagrangians with inverse powers of the electromagnetic field $F$ are interesting because the nonlinear electromagnetic effects might become important not only at early times in the cosmic evolution, but also at late times. Actually, models with Lagrangian density of the form \cite{novello-prd-2004} 

\bea {\cal L}=-\frac{1}{4}F-\frac{\gamma}{F},\label{l-1/f}\eea may account for the late-time stage of accelerated expansion of magnetic universes. A combination of positive a negative powers of $F$ have been also considered in \cite{novello-cqg-2007} 

\bea {\cal L}=-\frac{1}{4}F-\frac{\gamma}{F}+\alpha F^2.\label{l-nled}\eea This toy model correctly describes the main stages of the cosmic evolution and is free of the cosmological big bang singularity: At early times the quadratic term $\propto F^2$ -- which is responsible for a non singular bounce -- dominates, while the Maxwell term $\propto -F$ dominates in the radiation era. The term $\propto F^{-1}$ dominates at late times causing the universe to accelerate. 

Even if the above -- very simplified -- models of nonlinear electrodynamics coupled to general relativistic cosmology, describe hypothetical systems reminiscent of the fields in the real world, these models comprise interesting dynamical behavior that is worthy of independent investigation. The question is, would any theoretically plausible NLED-based EM Lagrangians provide viable cosmological models? 

Looking for an answer to the above question will reveal that several well-known such models are to be rejected due to violations of several fundamental principles of physics. Actually, although experiment (observations) is the supreme judge in deciding whether a given physical theory is right or wrong, there are a few physical principles on which the fundamental theories of physics are grounded, which should be satisfied by any compelling model of our universe. Among them causality and classical stability play a special role. One may think of these basic principles as a kind of coarse filter for plausible theories, while experimental/observational testing represents the finest possible such filter.

In this regard, our goals in this paper will be: 

\begin{enumerate}

\item To test the non negativity of the square sound speed looking for possible instability against small perturbations of the background energy density,

\item to look for possible violations of causality associated with superluminal propagation of these small perturbations,\footnote{Causality and the light-cone structure in NLED has been analyzed in \cite{goulart}.} and

\item to seek for occurrence/absence of curvature singularities of sudden and big rip types, 

\end{enumerate} in the NLED theories given by (\ref{l-b-i}), (\ref{l-model}), (\ref{l-model'}), (\ref{l-f2}), (\ref{l-1/f}), and (\ref{l-nled}), respectively. Additionally we shall check whether the sufficient conditions for the bounce are met by the different models. 

These simple tests can give an invaluable insight into the NLED-based models. As a matter of fact our study will show that the cosmological scenarios which are based in the NLED Lagrangians (\ref{l-b-i}), (\ref{l-model}), (\ref{l-model'}), (\ref{l-f2}), (\ref{l-1/f}), and (\ref{l-nled}), either are plagued by curvature singularities of the sudden and/or big rip type, or are classically unstable against small perturbations of the cosmological background due to negative sign of the square sound speed, or both. Not to mention the causality issues that may arise as a consequence of superluminal propagation of such small perturbations of the background.

The paper has been organized in the following form. In the next section we expose the essentials of NLED coupled to general relativity. The basic cosmological equations are discussed in section \ref{sec-cosmo}. Section \ref{sec-cs2} is dedicated to discussing on the stability and causality issues which are associated with violations of the bounds $0\leq c_s^2\leq 1$ on the square sound speed. The NLED-based cosmological models which are depicted by the Lagrangian densities (\ref{l-f2}), (\ref{l-1/f}), and (\ref{l-nled}) are studied in details in sections \ref{sec-f2}, \ref{sec-1/f}, and \ref{sec-nled}, respectively. Meanwhile the more complex BI model (\ref{l-b-i}) and its modifications (\ref{l-model}), and (\ref{l-model'}), are investigated in section \ref{sec-b-i}. The discussion of the results together with brief conclusions will be provided in the final section \ref{sec-disc}.

\section{Nonlinear Electrodynamics coupled to General Relativity}\label{sec-defi}

The four-dimensional (4D) Einstein-Hilbert action of gravity coupled to NLED is given by

\be S=\int d^4x\sqrt{-g}\left[ R+{\cal L}_\text{m}+{\cal L}(F,G)\right],\label{action}\ee where $R$ is the curvature scalar, ${\cal L}_\text{m}$ -- the background perfect fluid's Lagrangian density, and ${\cal L}(F,G)$ is the gauge-invariant electromagnetic (EM) Lagrangian density which is a function of the electromagnetic invariants $F$, defined in Eq. (\ref{f-inv}), and $$G\equiv\frac{1}{2}\,\epsilon_{\alpha\beta\mu\nu}F^{\alpha\beta}F_{\mu\nu}=-4{\bf E}\cdot{\bf B}.$$ 

Standard (linear) Maxwell electrodynamics is given by the Lagrangian ${\cal L}=-F/4$. The corresponding field equations can be derived from the action (\ref{action}) by performing variations with respect to the spacetime metric $g_{\mu\nu}$, to obtain: $$G_{\mu\nu}=T_{\mu\nu}^\text{m}+T_{\mu\nu}^{EM}\;,$$ where 

\bea &&T_{\mu\nu}^\text{m}=\left(\rho_\text{m}+p_\text{m}\right) u_{\mu}u_{\nu}-p_\text{m} g_{\mu\nu},\nonumber\\
&&T_{\mu\nu}^\textsc{EM}=g_{\mu\nu}\,\left[{\cal L}(F)-G {\cal L}_G\right]-4F_{\mu\alpha}F_\nu^{\;\;\alpha}\,{\cal L}_F,\label{em_t}\eea with $\rho_\text{m}=\rho_\text{m}(t)$, $p_\text{m}=p_\text{m}(t)$ -- the energy density and barotropic pressure of the background fluid, respectively, while ${\cal L}_F\equiv d{\cal L}/dF$, ${\cal L}_{FF}\equiv d^2{\cal L}/dF^2$, etc. Variation with respect to the components of the electromagnetic potential $A_\mu$ yields to the electromagnetic field equations\footnote{Here the comma denotes partial derivative in respect to the spacetime coordinates while the semicolon denotes covariant derivative instead.} 

\be \left(F^{\mu\nu}\,{\cal L}_F+\frac{1}{2}\epsilon^{\alpha\beta\mu\nu}F_{\alpha\beta}{\cal L}_G\right)_{;\mu}=0.\label{em-field}\ee

Since the observations have shown that the current universe is very close to a spatially flat geometry \cite{flat-1}, a result which is quite natural within primordial inflation scenarios \cite{flat-2}, in this paper we shall consider a homogeneous and isotropic Friedmann-Robertson-Walker (FRW) universe with flat spatial sections, which is described by the metric $$ds^{2}=dt^{2}-a(t)^2\delta_{ij}dx^idx^j,$$ where $a(t)$ is the cosmological scale factor, and the Latin indexes run over three-space. Since the spatial sections of the FRW spacetime are isotropic, the EM fields can be compatible with such a universe only if an averaging procedure is performed.\footnote{In particular, the energy density and the pressure of the NLED field should be evaluated by averaging over volume.} Following the standard approach \cite{tolman} (for details see also \cite{klippert, novello-prd-2004, novello-cqg-2007} and references therein) we define the volumetric spatial average of a quantity $X$ at the time $t$ by: 

\be \overline{X}\equiv \lim_{V\rightarrow V_{0}}\frac{1}{V}\int d^3x \sqrt{-g}\;X,\ee where $V=\int d^3x\sqrt{-g}$ and $V_{0}$ is a sufficiently large time-dependent three-volume. Besides, for the electromagnetic field to act as a source for the FRW model we need to impose that\footnote{The averaging procedure is independent of the equations of the EM field so it can be safely applied in the NLED case \cite{klippert}.}

\bea &&\overline{E}_{i}=0,\;\overline{B}_{i}=0,\;\overline{E_{i}B_{j}}=0,\nonumber\\
&&\overline{E_{i}E_{j}}=-\frac{1}{3}E^{2}g_{ij},\;\;\;\overline{B_{i}B_{j}}=-\frac{1}{3}B^{2}g_{ij}.\label{average}\eea 

Additionally it has to be assumed that the electric and magnetic fields, being random fields, have coherent lengths that are much shorter than the cosmological horizon scales. Under these assumptions the energy-momentum tensor of the EM field -- associated with the Lagrangian density ${\cal L}={\cal L}(F,G)$ -- can be written in the form of the energy-momentum tensor for a perfect fluid:

\be T_{\mu\nu}^\textsc{EM}=\left(\rho_\textsc{EM}+p_\textsc{EM}\right) u_{\mu}u_{\nu}-p_\textsc{EM} g_{\mu\nu},\label{tem}\ee where

\bea &&\rho_\textsc{EM}=-{\cal L}+G {\cal L}_G-4{\cal L}_F E^{2},\nonumber\\
&&p_\textsc{EM}={\cal L}-G {\cal L}_G-\frac{4}{3}\left( 2B^{2}-E^{2}\right) {\cal L}_F,\nonumber\eea $E^2$ and $B^2$ being the averaged electric and magnetic fields squared, respectively. In this paper we shall restrict to the case of nonlinear theories defined by ${\cal L}={\cal L}(F)$ \cite{novello-cqg-2007}, so that 

\bea &&\rho_\textsc{EM}=-{\cal L}-4{\cal L}_F E^{2},\nonumber\\
&&p_\textsc{EM}={\cal L}-\frac{4}{3}\left( 2B^{2}-E^{2}\right){\cal L}_F.\label{rho-p}\eea 

In what follows, to simplify the analysis, we shall consider a flat FRW universe which is filled with a ''magnetic fluid'', i. e., the electric component (squared) $E^2$ will be assumed vanishing -- only the average of the magnetic field squared $B^2$ is non vanishing -- a case which is dubbed in the bibliography as the ``magnetic universe''. This case turns out to be relevant in cosmology as long as the averaged electric field $E$ is screened by the charged primordial plasma, while the magnetic field lines are frozen \cite{lemoine}. Even this simplified picture can give important physical insights.

\section{Cosmological equations}\label{sec-cosmo}

Here we shall consider magnetic universes which are driven by EM Lagrangian densities that depend on the invariant $F=2B^2$ only. In this much simpler case, the cosmological equations can be written in the following form:

\bea &&3H^{2}=\rho_\text{m}-{\cal L},\;\;2\dot{H}=-\gamma_\text{m}\rho_\text{m}+\frac{4}{3}F {\cal L}_F,\nonumber\\
&&\dot{\rho}_\text{m}+3H\gamma_\text{m}\rho_{\gamma}=0,\;\;\dot{F}+4HF=0,\label{feqs}\eea where $H={\dot a}/a$ is the Hubble parameter, $\gamma_\text{m}$ is the barotropic index of the background's perfect fluid [$p_\text{m}=\left(\gamma_\text{m}-1\right)\rho_\text{m}$], while 

\bea &&\rho_\textsc{EM}=\rho_\textsc{b}=-{\cal L},\;p_\textsc{EM}=p_\textsc{b}={\cal L}-\frac{4}{3}F {\cal L}_F,\nonumber\\
&&\Rightarrow\;\rho_\textsc{b}+p_\textsc{b}=-\frac{4}{3}F {\cal L}_F,\label{rhob-pb}\eea and, as already mentioned, we are considering $F=2B^2$. For simplicity, in the rest of this paper we shall assume a purely magnetic universe, i. e., the cosmic dynamics is fueled by the magnetic fluid alone:

\bea &&3H^{2}=\rho_\textsc{b},\;\;\dot{F}+4HF=0,\nonumber\\
&&2\dot{H}=-(\rho_\textsc{b}+p_\textsc{b})=\frac{4}{3}F {\cal L}_F,\label{feq}\eea where $\rho_\textsc{b}$ and $p_\textsc{b}$ are given in Eq. (\ref{rhob-pb}). 

An interesting aspect of several cosmological models based in NLED -- assuming a magnetic universe -- is that the cosmological scale factor $a=a(t)$ attains a minimum value during the course of the cosmic evolution. At this point a bounce occurs. At the bounce, since the scale factor is a minimum while $H$ changes sign (contraction turns into expansion), then $H=0$. In general, the sufficient conditions for a bounce are \cite{phys-rep-2008}:

\bea \dot a=0,\;\ddot a\geq 0\;\Leftrightarrow\;H=0,\;\dot H\geq 0,\label{bounce-c}\eea where the above quantities are evaluated at the bounce. Hence, it is customary to check whether the conditions (\ref{bounce-c}) for the bounce are met by a given cosmological model which is based in NLED.

Of particular interest for the bouncing cosmologies is the behavior of the background energy density. At the bounce at $t=t_b$, since $H(t_b)=0$, then the continuity equation yields that $\dot\rho_\textsc{b}(t_b)=0$, i. e., the energy density of the magnetic field is a critical value at the bounce. On the other hand, since $\rho_\textsc{b}=\rho_\textsc{b}(F(t))$, one has 

\bea \dot\rho_\textsc{b}=\frac{d\rho_\textsc{b}}{dF}\,\dot F,\;\ddot\rho_\textsc{b}=\frac{d^2\rho_\textsc{b}}{dF^2}\dot F^2+\frac{d\rho_\textsc{b}}{dF}\ddot F.\label{rho-extrema}\eea It follows that $\rho_\textsc{b}$ can have extrema with respect to both $F$ and the cosmic time $t$. In general, these not need to coincide. Actually, at $t_b$ the field invariant is a maximum $F(t_b)=F_b$, so that $\dot F(t_b)=0$, while $\ddot F(t_b)<0$. Suppose that $\rho_\textsc{b}$ is a maximum at some $F_c\neq F_b$. Two situations may arise:

\begin{enumerate}

\item $F_c<F_b$, where $$F_c=F(t^-_c)=F(t^+_c),\;t^-_c<t_b<t_c^+.$$ In this case, since $$\frac{d\rho_\textsc{b}}{dF}({F_c})=0,\;\frac{d^2\rho_\textsc{b}}{dF^2}(F_c)<0,$$ just prior to the bounce and just after it (at $t^-_c$ and $t^+_c$ respectively), the energy density of the magnetic field is a maximum as well (check Eq. (\ref{rho-extrema})): $$\dot\rho_\textsc{b}(t^\pm_c)=0,\;\ddot\rho_\textsc{b}(t^\pm_c)<0.$$ Meanwhile, at the bounce ($t=t_b$), since $d\rho_\textsc{b}/dF<0$, $\dot F=0$ and $\ddot F<0$, then $\ddot\rho_\textsc{b}(t_b)>0$, i. e., the energy density of the magnetic field is a local minimum.

\item $F_c>F_b$. In this case $d\rho_\textsc{b}/dF>0$, i. e., in respect to $F$, $\rho_\textsc{b}$ is a monotonic growing function in the interval $0\leq F\leq F_b<F_c$. Then the energy density of the magnetic field has only one extremum at $t=t_b$, where $\rho_\textsc{b}$ is a maximum: $$\dot\rho_\textsc{b}(t_b)=\frac{d\rho_\textsc{b}}{dF}\,\dot F(t_b)=0,\;\ddot\rho_\textsc{b}=\frac{d\rho_\textsc{b}}{dF}\ddot F(t_b)<0.$$

\end{enumerate}

\section{The square sound speed}\label{sec-cs2}

Another quantity of cosmological importance is the adiabatic square sound speed which, for the cases of interest in this paper, can be written as

\bea c_s^2:=\frac{dp_\textsc{b}}{d\rho_\textsc{b}}=\frac{dp_\textsc{b}/dF}{d\rho_\textsc{b}/dF}=\frac{1}{3}+\frac{4{\cal L}_{FF}}{3{\cal L}_F}\,F.\label{speed-sound}\eea 

If consider small perturbations of the background energy density $\rho_\textsc{b}(t,{\bf x})=\rho_\textsc{b}(t)+\delta\rho_\textsc{b}(t,{\bf x})$, the conservation of energy-momentum $T^{\mu\nu}_{\;\;;\nu}=0$, leads to the wave equation \cite{peebles-ratra}: $\delta\ddot\rho=c_s^2\nabla^2\delta\rho$, which solution for positive $c_s^2>0$ is $$\delta\rho_\textsc{b}=\delta\rho_\textsc{b0}\exp(-i\omega t+i{\bf k}\cdot{\bf x}),$$ while, for negative $c_s^2<0$, it is $$\delta\rho_\textsc{b}=\delta\rho_\textsc{b0}\exp(\omega t+i{\bf k}\cdot{\bf x}).$$ Here $\omega=k\,c_s$, where $k=2\pi/\lambda$ is the wave number of the perturbation ($a/k$ is the physical wavelength of the perturbation). In the case when $c_s^2<0$, the energy density perturbations uncontrollably grow resulting in a classical instability of the cosmological model. The increment of instability is inversely proportional to the wavelength of the perturbations, and the models where $c_s^2<0$, are violently unstable, so that these should be rejected \cite{chinos}. In consequence, it is very important to check the sign of the square sound speed for the different cosmological models.

\subsection{Causality and the speed of sound}

Even if $c_s^2$ is a positive quantity, a causality issue may arise whenever the square sound speed is greater than the local speed of light (for a critical review on this issue see \cite{roy}). As a matter of fact, it is usually assumed that $c_s\leq 1$, while the complementary bound $c_s>1$ is used as a criterion for rejecting theories \cite{hawking-ellis, wald}. In particular, low-energy effective field theories -- even when these are based in Lorentz-invariant Lagrangians -- have been rejected if they admit superluminal fluctuations \cite{nima}. Notwithstanding, there can be found arguments which challenge the most widespread viewpoint (see, for instance, Ref. \cite{caldwell}). A related illustration can be found in the reference \cite{novello-cqg-2007}, where it was argued that nonlinear photons do not move on the light cone of the background metric $g_{\mu\nu}$, but instead, these follow the null rays of the effective metric 

\bea g^\text{eff}_{\mu\nu}={\cal L}_F g_{\mu\nu}-4{\cal L}_{FF} F_\mu^{\;\;\sigma}F_{\sigma\nu},\label{eff-metric}\eea so that, according to \cite{novello-cqg-2007}, this fact may ``introduce a new look into causality''. It follows that the signature of the effective metric $g^\text{eff}_{\mu\nu}$ is undefined. Take as an example, the Lagrangian density (\ref{l-f2}). In this case $0\leq F\leq 1/4\alpha$. The energy density of the magnetic field $\rho_\textsc{b}=F(1-4\alpha F)/4$, is a maximum at $F_c=1/8\alpha$ ($\rho^\text{max}_\textsc{b}=1/64\alpha$). Since ${\cal L}_F=(8\alpha F-1)/4$, then $$g^\text{eff}_{\mu\nu}=\left(\frac{8\alpha F-1}{4}\right) g_{\mu\nu}-8\alpha F_\mu^{\;\;\sigma}F_{\sigma\nu}.$$ In the FRW magnetic universe, for the $(0,0)$-component of the metric one has $g^\text{eff}_{00}=(8\alpha F-1)/4$. Hence, at the maximum of the energy density ($F=1/8\alpha$) a signature change occurs. While for $1/8\alpha<F\leq 1/4\alpha$ the signature of the effective metric $g^\text{eff}_{\mu\nu}$ coincides with that of the gravitational metric $g_{\mu\nu}$, for $0\leq F<1/8\alpha$ these have different signature. The signature change of the effective metric is awful for causality to be satisfied for all $0\leq F\leq 1/4\alpha$ if one considers $g^\text{eff}_{\mu\nu}$ to be the ``arbiter of causality'' instead of the gravitational metric.

In general, as discussed in \cite{roy}, the sound cones for any given fluid can be represented by an appropriate effective (hyperbolic) metric tensor $$g^\text{eff}_{\mu\nu}=g_{\mu\nu}+\frac{1-c_s^2}{c_s^2}\,h_{\mu\nu}=\frac{1}{c_s^2}\left[g_{\mu\nu}+(1-c_s^2)u_\mu u_\nu\right],$$ where $h_{\mu\nu}=g_{\mu\nu}+u_\mu u_\nu$ projects into the rest space at each event, and $u_\mu$ is the matter 4-velocity. When $c_s^2>1$ the sound cones will lie outside the light cones. In this case for both metrics $g_{\mu\nu}$ and $g^\text{eff}_{\mu\nu}$, the interior of the light cones consists of timelike vectors. According to the unorthodox view point, in the case when the sound speed is superluminal, one can safely redefine the physical metric to be $g^\text{eff}_{\mu\nu}$, and there will be no problem with causality. This view point is totally wrong since, following the line of reasoning of \cite{roy}, one could find that the sound metric $g^\text{eff}_{\mu\nu}$ is in some places superluminal and in others subluminal. Hence, at least at some events and in some directions, part of the light cone could lie outside the sound cone. As a consequence gravitons -- for instance -- could propagate acausally relative to the sound metric. As a matter of fact, if one wants to preserve the principle that the effects of gravity are encoded in the spacetime curvature, then one may not abandon the spacetime metric $g_{\mu\nu}$ as the arbiter of causality \cite{roy}.

\subsection{Bounds on $c_s^2$}

In this paper, following the most widespread point of view, we shall consider $c_s^2>1$ as a criterion for rejecting a given cosmological model, as long as causality is violated in it. Our choice is based on solid long standing arguments which are comprised in well-tested and theoretically beautiful physical theories. Hence, a given cosmological model which is intended to describe the present universe, has to meet the following bounds on the speed at which small perturbations of the background energy density propagate:

\bea 0\leq c_s^2\leq 1.\label{bounds}\eea This is one of the aspects of the NLED-based theories (\ref{l-b-i}-\ref{l-nled}) which we shall meticulously check.


\begin{figure*}[t!]\begin{center}
\includegraphics[width=7cm,height=6cm]{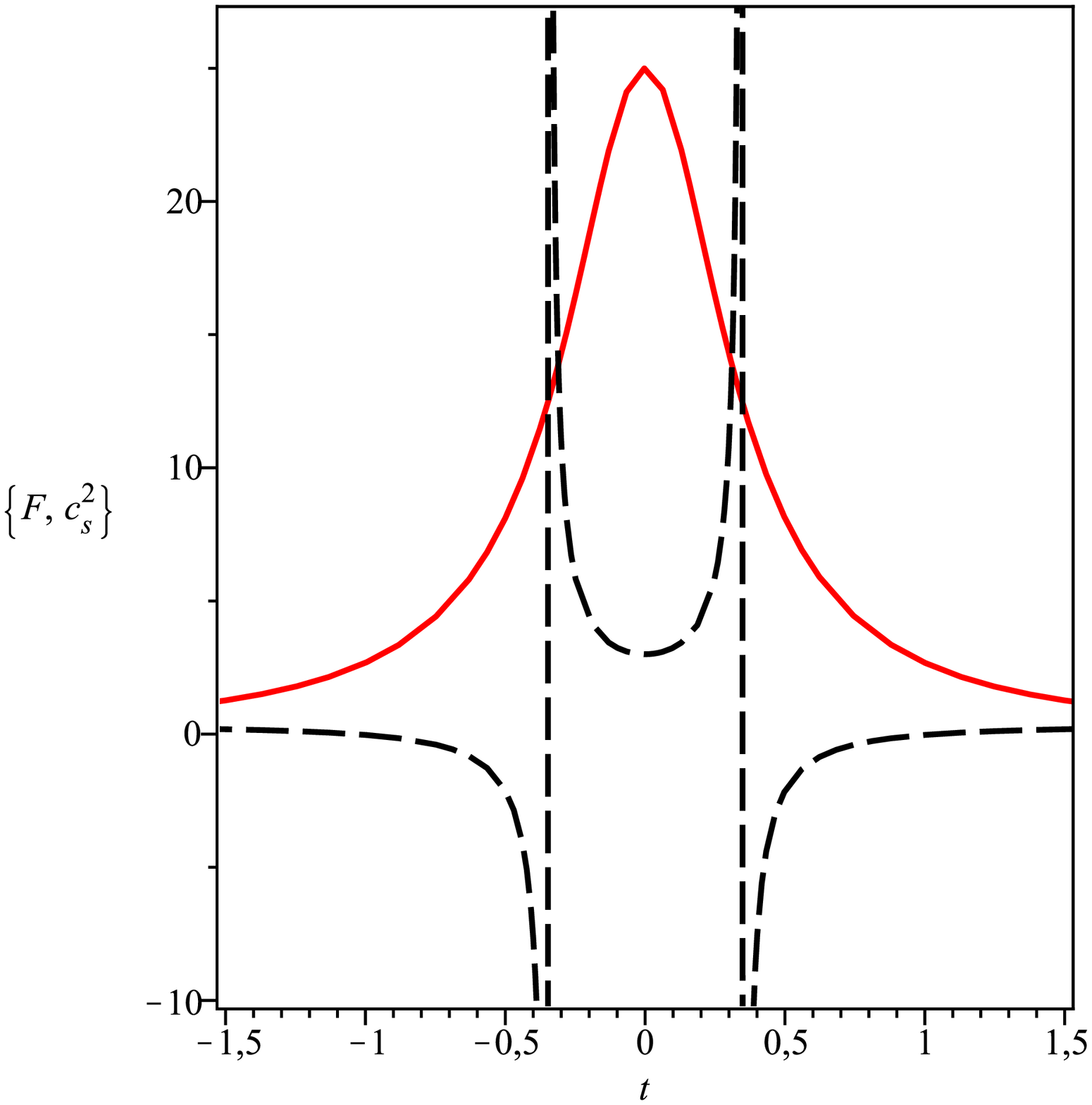}
\includegraphics[width=7cm,height=6cm]{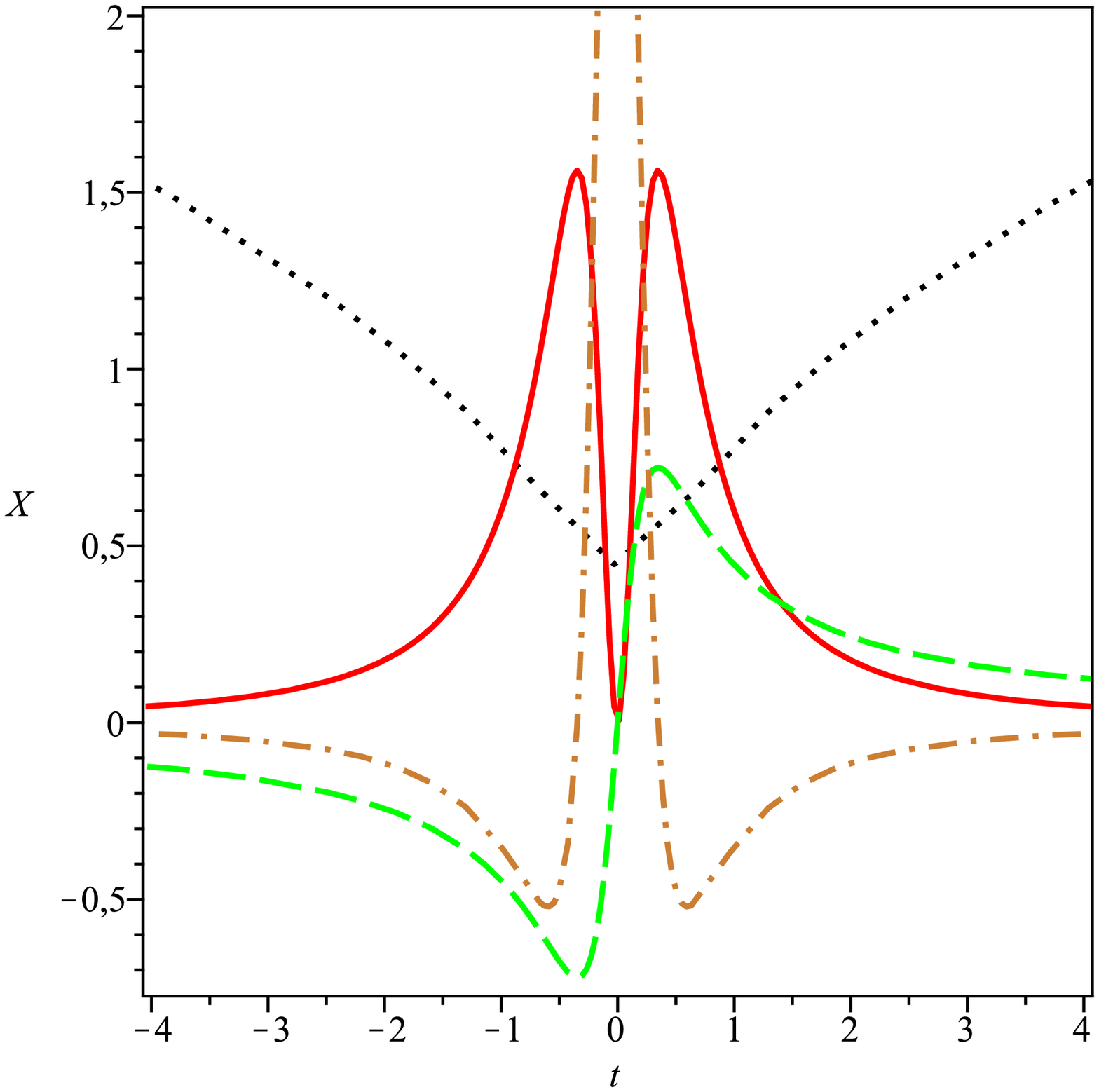}
\end{center}\vspace{0.3in}
\caption{In the left-hand panel a plot of $F$ -- solid curve -- and of the square sound speed $c_s^2$ -- dashed curve -- vs the cosmic time $t$, is shown for an arbitrarily chosen $\alpha=0.01$, for the model (\ref{l-f2}) [${\cal L}=-F/4+\alpha F^2$]. For the same theory a plot of $X=\{\rho_\textsc{b}(t)-\text{solid},H(t)-\text{dash},\dot H(t)-\text{dashdot},a(t)-\text{dots}\}$ is shown in the right-hand panel. Occurrence of a bounce at $t=0$ is evident from this latter figure.}\label{fig1}\end{figure*}


\section{Second order NLED: bouncing magnetic universe?}\label{sec-f2}

We start our investigation with the toy model generalization of Maxwell EM Lagrangian which contains terms up to second order in the field invariant $F$, and was proposed in Ref. \cite{klippert} (see also \cite{novello-cqg-2007}). This model is based in the Lagrangian density (\ref{l-f2}). The corresponding associated energy density of the magnetic field

\bea \rho_\textsc{b}=F(1-4\alpha F)/4,\label{rho-f2}\eea is non negative in the interval $0\leq F\leq 1/4\alpha$. At $F=1/8\alpha$, it is a maximum $\rho^\text{max}_\textsc{b}=1/64\alpha$. The corresponding parametric pressure

\bea p_\textsc{b}=F(1-20\alpha F)/12,\label{p-f2}\eea is a negative quantity for $F>1/20\alpha$. 

Taking into account (\ref{rho-f2}), the Friedmann equation $H=\sqrt{\rho_\textsc{b}/3}$, and the continuity equation $\dot F=-4H F$, one can find that

\bea &&a(t)=\left(F_0/3\right)^{1/4}\left(t^2+12\alpha\right)^{1/4},\nonumber\\
&&F(t)=\frac{3}{t^2+12\alpha},\;\rho_\textsc{b}=\frac{3t^2}{4(t^2+12\alpha)^2}.\label{sol-f2}\eea  As seen, the scale factor is a minimum at $t=0$ (dotted curve in the right-hand panel (RHP) of FIG. \ref{fig1}) $$a_\text{min}=a(0)=(4\alpha F_0)^{1/4},$$ i. e., the magnetic universe in this NLED model is in a stage of contraction until it reaches a minimum size at $t=0$ (time of the bounce), and then starts expanding. In terms of the cosmological time, the energy density of the magnetic field (solid curve in the RHP of FIG. \ref{fig1}) is a vanishing minimum $\rho^\text{min}_\textsc{b}=0$ at the bounce ($t=0$), while it is a maximum $\rho^\text{max}_\textsc{b}=1/64\alpha$ at $t=\pm 2\sqrt{3\alpha}$. Notice in between that at $t=0$ the field invariant $F=2B^2$ is a maximum. This means that at the bounce the nonlinear effects are maximal (of the same order as the Maxwellian effects), but contribute a negative fraction to the magnetic field energy density, so that these cancel the contribution to the field energy density coming from standard Maxwell EM theory. The vanishing minimum of the energy density -- together with positivity of $\dot H$ -- is one of the sufficient conditions for the occurrence of a bounce at $t=0$. 

The bad news for this model comes from the behavior of the square sound speed $c_s^2$ during the cosmic history. Actually, in this case $c_s^2$ is given by

\bea c_s^2=\frac{1}{3}\left(\frac{1-40\alpha F}{1-8\alpha F}\right)=\frac{1}{3}\left(\frac{t^2-108\alpha}{t^2-12\alpha}\right).\label{cs2-f2}\eea It is a non negative quantity whenever either $0\leq F\leq 1/40\alpha$, or $1/8\alpha<F\leq 1/4\alpha$. However, in the interval $1/40\alpha<F<1/8\alpha$, $c_s^2$ is negative. At $F=1/8\alpha$ it has a vertical asymptote. In terms of the cosmic time the square sound speed -- dashed curve in the left-hand panel (LHP) of FIG. \ref{fig1} -- has vertical asymptotes at $t_\pm=\pm 2\sqrt{3\alpha}$. It is a negative quantity whenever $$-6\sqrt{3\alpha}<t<-2\sqrt{3\alpha},$$ during the contracting phase, and $$2\sqrt{3\alpha}<t<6\sqrt{3\alpha},$$ in the expanding stage of the cosmic evolution. This means that during the contraction, just before the moment $t_-=-2\sqrt{3\alpha}$, and after the moment $t_+=2\sqrt{3\alpha}$ in the expanding phase, there are intervals of cosmic time $\Delta t=4\sqrt{3\alpha}$, where the square sound speed is negative, signaling an insurmountable instability of the cosmic evolution. Besides, since $$\dot c_s^2=\frac{192 t}{3(t^2-12\alpha)^2},$$ at the bounce ($t=0$) the square sound speed is a minimum given that $\ddot c_s^2(0)=192/432\alpha>0$. At the minimum $c_s^2(0)=c^2_{s,b}=108/36>1$, i. e., at the bounce -- where the non linear EM effects dominate the cosmic evolution -- the superluminal propagation of the small fluctuations of the background energy density, violates causality. 

The details of this can be seen from FIG. \ref{fig1}, where the plot of the main cosmological parameters [$F$, $a$, $c_s^2$, $\rho_\textsc{b}$, $H$, $\dot H$] vs $t$ is shown for an arbitrarily chosen value of the free parameter $\alpha=0.01$. It is seen that, as the cosmic evolution proceeds from the past (negative $t$-s), the universe transits from a stage of contraction -- through a bounce at $t=0$ -- into a stage of cosmic expansion. Besides, as one approaches the bounce from the past, the square sound speed $c_s^2$ gets increasingly negative [$|c_s^2|\rightarrow\infty$, $c_s^2<0$], and at $t=-2\sqrt{3\alpha}$ -- when the energy density of the magnetic field becomes a maximum right before the bounce -- $c_s^2$ approaches to a vertical asymptote $$t\rightarrow-2\sqrt{3\alpha}\;\Rightarrow\;c_s^2\rightarrow-\infty.$$ As one leaves behind the bounce at $t=2\sqrt{3\alpha}$ -- where the second maximum of $\rho_\textsc{b}$ arises -- there is a second asymptote to the right of which another stage of the cosmic evolution with negative $c_s^2<0$ occurs. 

The intervals of cosmic time $\Delta t=4\sqrt{3\alpha}$, before the bounce at $t=0$ and after it, are critical and decide the fate of this model. Actually, since to the left of $t_-=-2\sqrt{3\alpha}$ and to the right of $t_+=2\sqrt{3\alpha}$, $|c_s^2|$ is very large and negative, these periods of cosmological time are characterized by insurmountable instabilities against small fluctuations of the background. This means that, even if this is a regular cosmological model which is free of curvature singularities,\footnote{The addition of other kinds of matter --including their ultrarelativistic states -- does not modify the regularity of the corresponding solutions \cite{klippert}.} as the universe transits from the past -- through the bounce -- into the future, it will not survive contraction behind $t_-$. For the same reason the cosmic evolution predicted by this model will not survive expansion after the bounce past $t_+$. This is not to mention the causality issue that arises due to the fact that the speed of sound exceeds the speed of light [$c_s^2>1$] in between the asymptotes.


\begin{figure*}[t!]\begin{center}
\includegraphics[width=7cm,height=6cm]{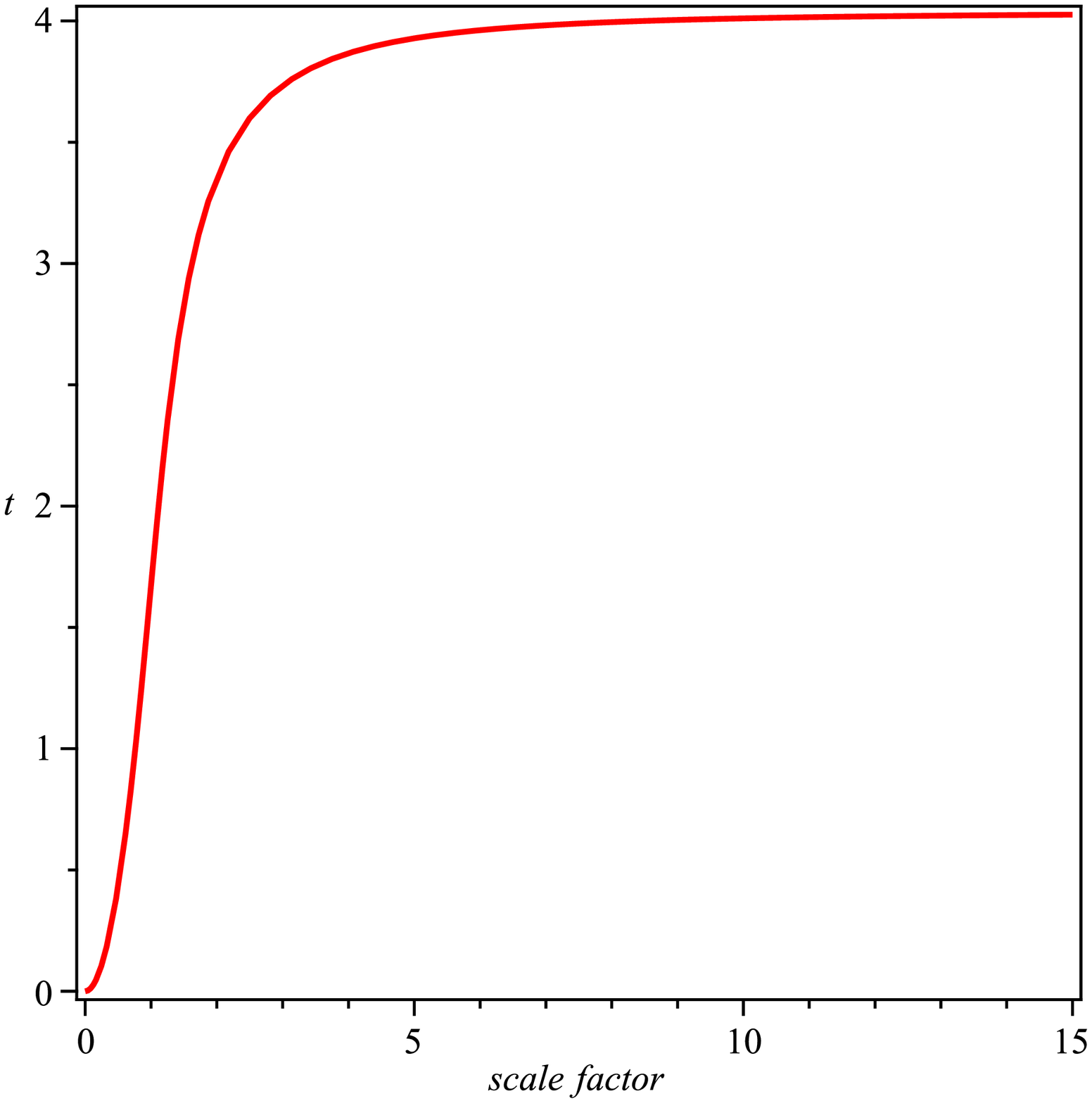}
\includegraphics[width=7cm,height=6cm]{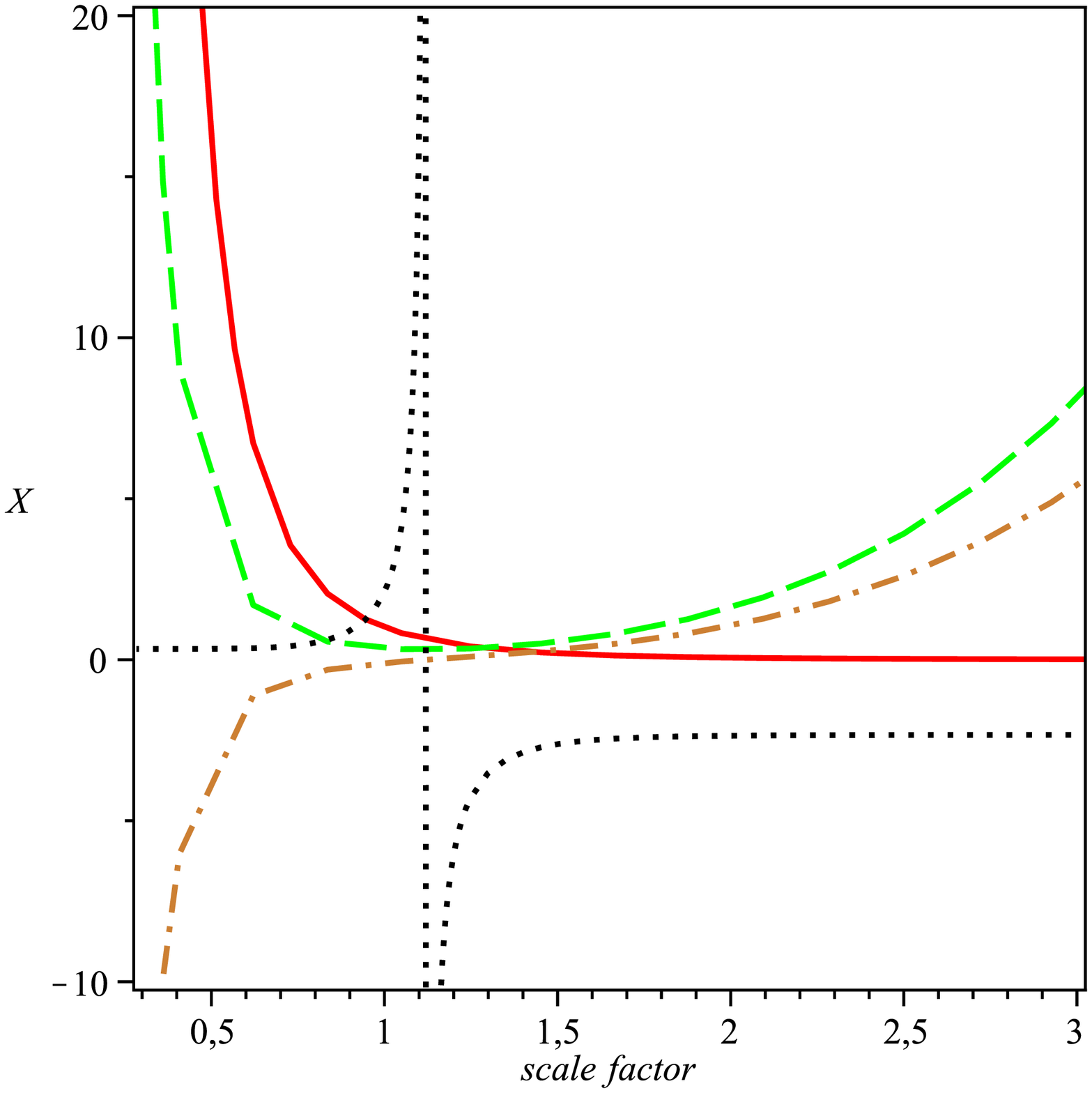}
\end{center}\vspace{0.3in}
\caption{Plots for the model (\ref{l-1/f}) [${\cal L}=-F/4-\gamma/F$] for an arbitrarily chosen value of the parameter $\gamma=0.1$. In the left-hand panel the plot of the cosmic time $t$ vs the scale factor $a$ is shown, while in the right-hand panel we have plotted $X=\{F(a)-\text{solid},\rho_\textsc{b}(a)-\text{dash},\dot H(a)-\text{dashdot},c_s^2 - \text{dots}\}$.}\label{fig2}\end{figure*}

\begin{figure}[t!]\begin{center}
\includegraphics[width=7cm,height=6cm]{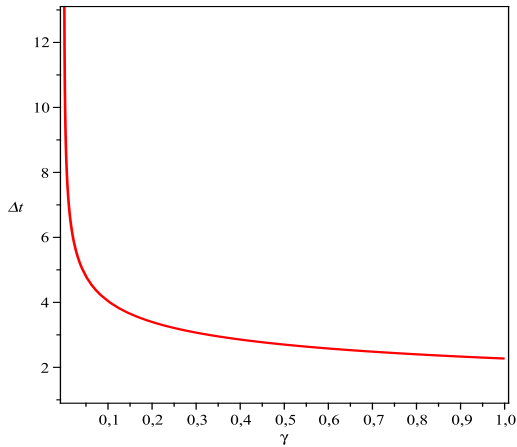}
\end{center}\vspace{0.3in}
\caption{Plot of the time duration of the cosmic history $\Delta t$ vs the free parameter $\gamma$, for the model (\ref{l-1/f}) [${\cal L}=-F/4-\gamma/F$]. Only the parameter interval $\gamma\in[0.001,1]$ is shown.}\label{fig3}\end{figure}


\section{NLED-driven accelerated expansion?}\label{sec-1/f}

In the reference \cite{novello-prd-2004}, in order to take account of the late time accelerated stage of the cosmic expansion -- without invoking the cosmological constant, unobserved scalar fields or modifications of general relativity -- the authors proposed the gauge invariant Lagrangian density (\ref{l-1/f}) which, besides the standard Maxwell term, contains a negative power of the field invariant $F$. At high values of $F$ the dynamics will be that of Maxwell -- driving a standard radiation dominated stage of the cosmic evolution -- plus corrections which are regulated by the parameter $\gamma$, while at small values of $F$ it is the $1/F$ term which dominates, driving the late time accelerated stage of the cosmic expansion \cite{novello-prd-2004} (see also \cite{novello-cqg-2007}). We want to notice at this point that -- as appropriately discussed in \cite{novello-prd-2004} -- this model lacks a standard Maxwellian weak field limit, but other more profound problems will be revealed soon.

The energy density associated with the Lagrangian density (\ref{l-1/f}) is given by

\bea \rho_\textsc{b}=\frac{F^2+4\gamma}{4F},\label{rho-1/f}\eea is always a positive quantity, i. e., the EM invariant $F$ can take values in the whole (non negative) real line ($0\leq F<\infty$). The energy density of the EM field is a minimum $\rho^\text{min}_\textsc{b}=\sqrt\gamma$ at $F=2\sqrt\gamma$. The parametric pressure

\bea p_\textsc{b}=\frac{F^2-28\gamma}{12 F},\label{p-1/f}\eea is negative for $0\leq F<\sqrt{28\gamma}$. 

The square sound speed is given by

\bea c_s^2=\frac{F^2+28\gamma}{3(F^2-4\gamma)}.\label{cs2-1/f}\eea It is a negative quantity for $0\leq F<2\sqrt\gamma$, and at $F=2\sqrt\gamma$ -- where $\rho_\textsc{b}$ is a minimum -- it has a vertical asymptote.

Taking into account the continuity equation $\dot F=-4H F$ $\Rightarrow$ $F=F_0 a^{-4}$, the Friedmann equation can be integrated to obtain:

\bea t=t(a)=\sqrt\frac{3}{2i\sqrt\gamma}\,F\left(\sqrt\frac{2i\sqrt\gamma}{F_0}\,a^2,i\right),\label{sol-1/f}\eea where $F(z,k)$ is the elliptic integral of the first kind\footnote{Do not confound with the EM invariant $F$.} $$F(z,k)=\int_0^z\frac{d\xi}{\sqrt{1-\xi^2}\sqrt{1-k^2\xi^2}},$$ and $i$ is the imaginary unit. In principle the equation (\ref{sol-1/f}) can be inverted to get $a=a(t)$, but in practice this is a very difficult task and only numeric investigation may help. Nevertheless, as seen from the LHP of FIG. \ref{fig2}, the scale factor is a monotonically increasing -- smooth and continuous -- function of the cosmic time. Hence, in this subsection instead of the cosmic time $t$ we can use the scale factor as a time ordering variable, i. e., we will study the dynamics of the field variables $F$, $\rho_\textsc{b}$, $\dot H$, $c_s^2$, etc., in respect to the scale factor $a$:

\bea &&\rho_\textsc{b}=\frac{F_0^2+4\gamma a^8}{4F_0 a^4},\;p_\textsc{b}=\frac{F_0^2-28\gamma a^8}{12F_0 a^4},\nonumber\\
&&\dot H=-\frac{F_0^2-4\gamma a^8}{6F_0 a^4},\;c_s^2=\frac{F_0^2+28\gamma a^8}{3(F_0^2-4\gamma a^8)}.\label{sol-a-1/f}\eea This will amount to a great simplification of the analysis. 

Another relevant aspect of the plot $t(a)$ -- which is evident from the LHP of FIG. \ref{fig2} -- is the fact that there is a finite time duration $\Delta t$ of the cosmic expansion in the model (\ref{l-1/f}). In the FIG. \ref{fig3} a plot of $\Delta t$ vs the free parameter $\gamma$ is shown. As clearly seen, the smaller $\gamma$ is, the larger $\Delta t$.\footnote{For the chosen value $\gamma=0.1$, $\Delta t=4.038$, while for other values one would have: $(\gamma,\Delta t)\;\Rightarrow$ $(1,2.2707)$, $(0.1,4.038)$, $(0.01,7.1808)$, $(0.001,12.7694)$, $(0.0001,22.7076)$, etc. For $\gamma=0$ one would have an infinite time duration of the cosmic history [$\Delta t\rightarrow\infty$] as it should be for a universe fueled by Maxwell EM fields.} The finite time duration of the cosmic history -- into the future -- signals a big rip-type singularity since the scale factor blows up within a finite time interval \cite{big-rip}. As seen also from FIG. \ref{fig2}, the energy density of the magnetic field is a minimum [$\rho^\text{min}_\textsc{b}=\sqrt\gamma$] at the value of the scale factor $a_*=(F_0/2\sqrt\gamma)^{1/4}$. Notice also that as the scale factor grows up above $a_*$, the energy density of the field starts increasing in such a way that $$a\rightarrow\infty\;\Rightarrow\;\rho_\textsc{b}\rightarrow\infty.$$ In other words, in the NLED model (\ref{l-1/f}) the magnetic field behaves like a phantom field, but without actual phantom matter. Within the finite time taken by the universe to go from a size $\sim a_*$ to an infinite size, the energy density of the magnetic field grows up from $\rho^\text{min}_\textsc{b}$ to an infinite density universe, while $\dot H$ goes from negative values into infinite positive values. This is why the fate of the cosmic evolution in this model is a catastrophic big rip singularity. Summarizing: in the model of Ref. \cite{novello-prd-2004} -- see also \cite{novello-cqg-2007} -- there occur two curvature singularities: (i) the initial big bang singularity,\footnote{At $a=0$, since there is a big bang singularity, $\rho_\textsc{b}$, as well as $H$, $\dot H$, and $p_\textsc{b}$, blow up.} and (ii) the final big rip singularity. The entire cosmic history interpolates between these two curvature singularities. The above mentioned features have been appropriately discussed in Ref. \cite{novello-prd-2004}.

There are, however, other troublesome aspects of this model which were not discussed in \cite{novello-prd-2004}. These are related with the classical stability and with causality. Actually, as seen from the RHP of FIG. \ref{fig2}, at $a_*=(F_0/2\sqrt\gamma)^{1/4}$ where $\rho_\textsc{b}$ is a minimum, there is a vertical asymptote of $c_s^2(a)$ where the square sound speed blows up. To the right of the asymptote [$a(t)>a_*$] the square sound speed is a negative quantity. Hence, starting at the universe size where the term $1/F$ dominates -- i. e., at the minimum of $\rho_\textsc{b}$ -- and up to the end of the cosmic history, $c_s^2<0$ in this model. This means that a violent instability against small fluctuations of the background is developed due to the term $1/F$, which causes that the model is really unable to take account of the accelerated phase of the cosmic expansion. Additionally, as seen in the RHP of FIG. \ref{fig2}, there is an $F$-interval to the left of the asymptote at $a_*$, where the square sound speed generously exceeds the local speed of light squared: $$\lim_{a\rightarrow a_*^-}c_s^2=\infty.$$ This raises a not less important causality issue due to superluminal propagation of the small fluctuations of the background energy density around $\rho_\textsc{b}=\rho^\text{min}_\textsc{b}$.


\begin{figure*}[t!]\begin{center}
\includegraphics[width=5.5cm,height=5cm]{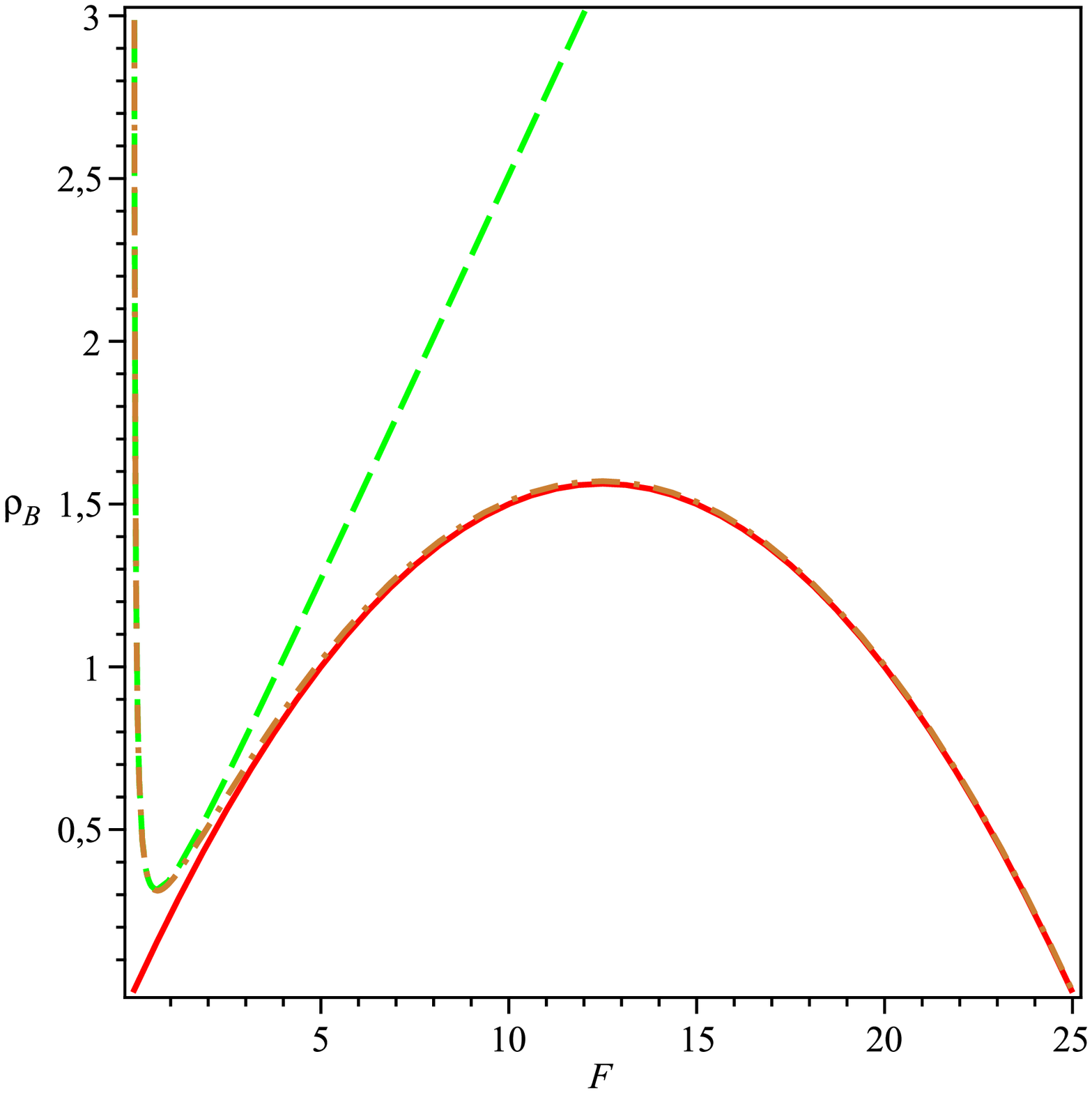}
\includegraphics[width=5.5cm,height=5cm]{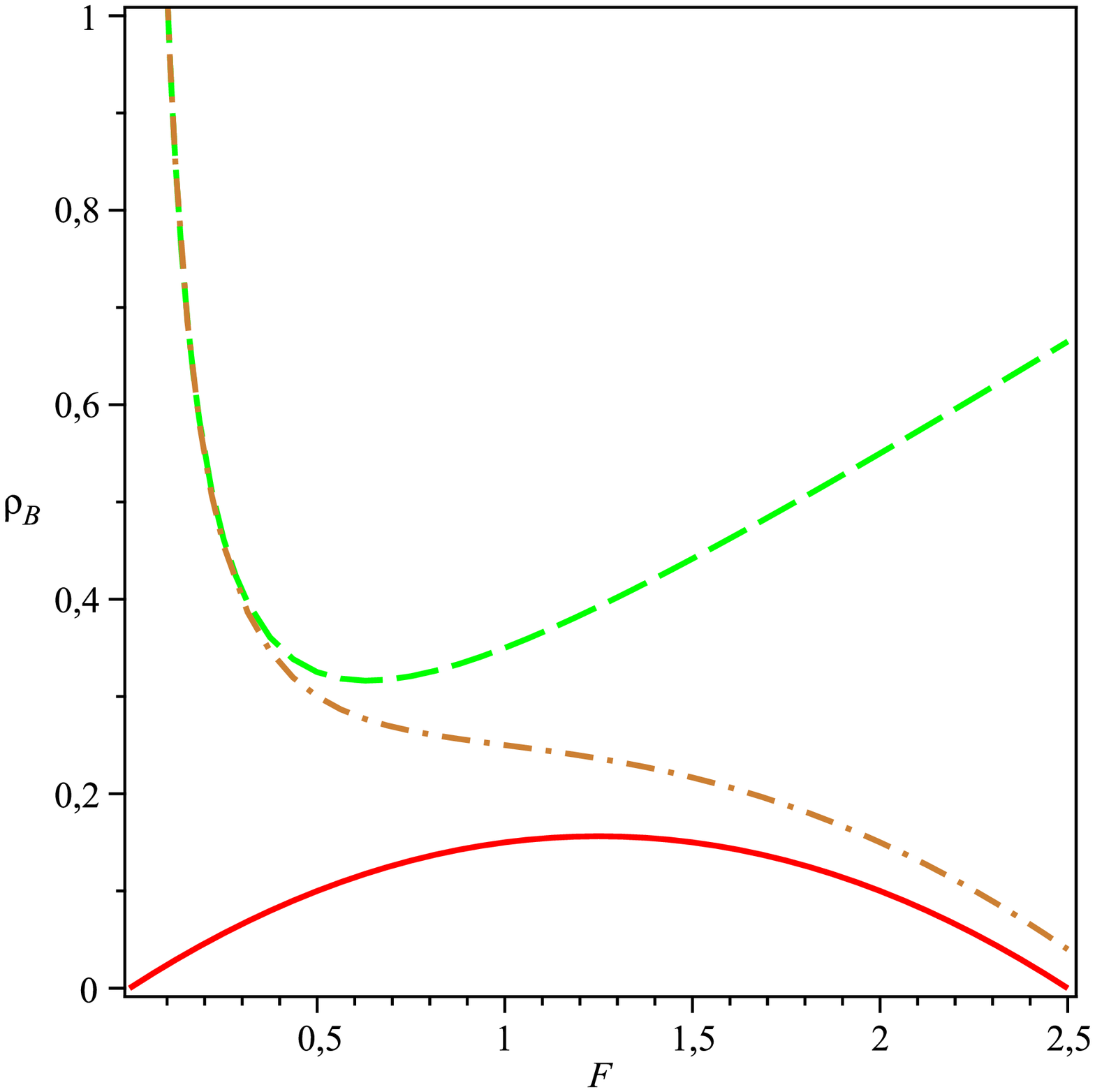}
\includegraphics[width=5.5cm,height=5cm]{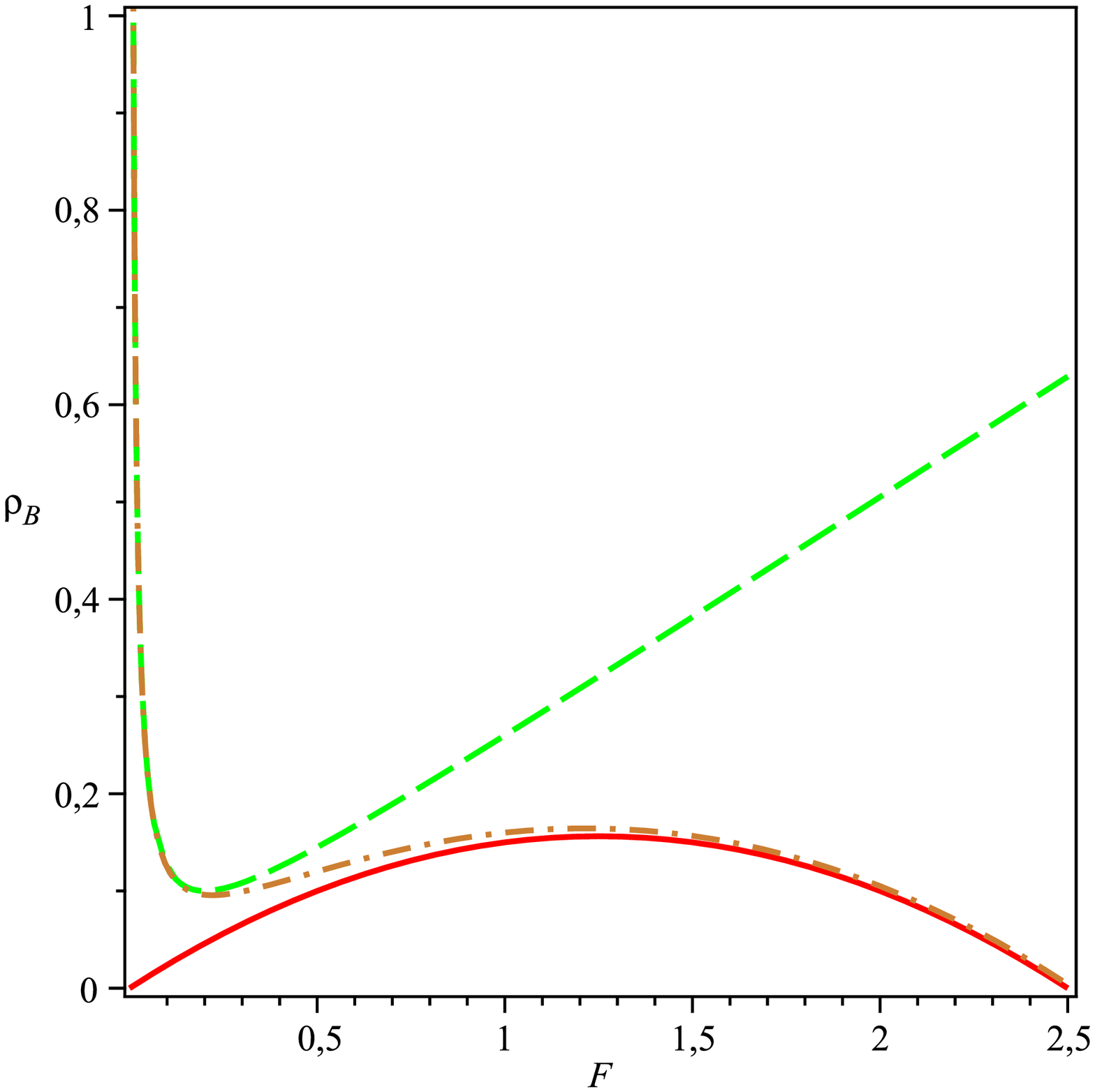}
\includegraphics[width=5.5cm,height=5cm]{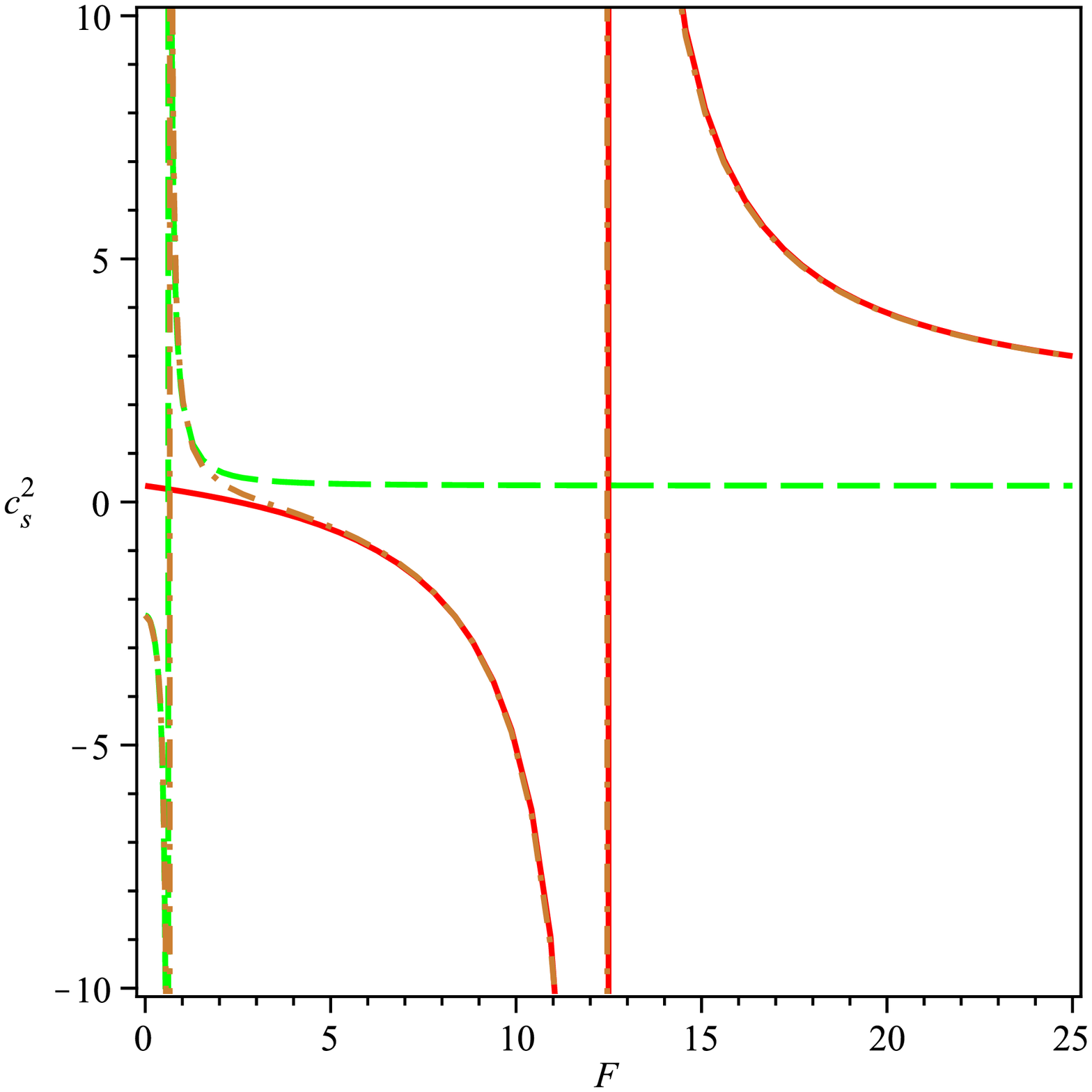}
\includegraphics[width=5.5cm,height=5cm]{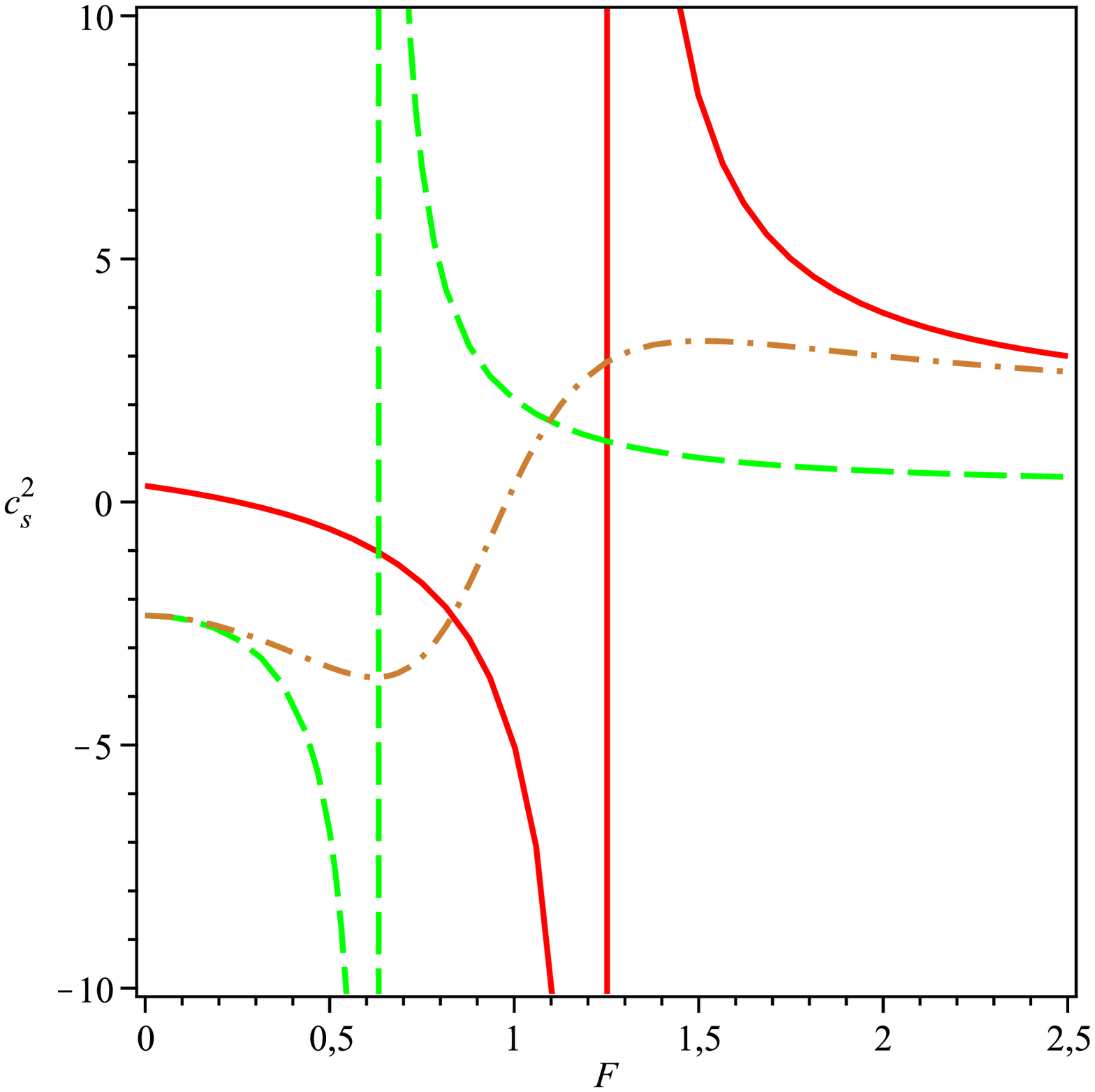}
\includegraphics[width=5.5cm,height=5cm]{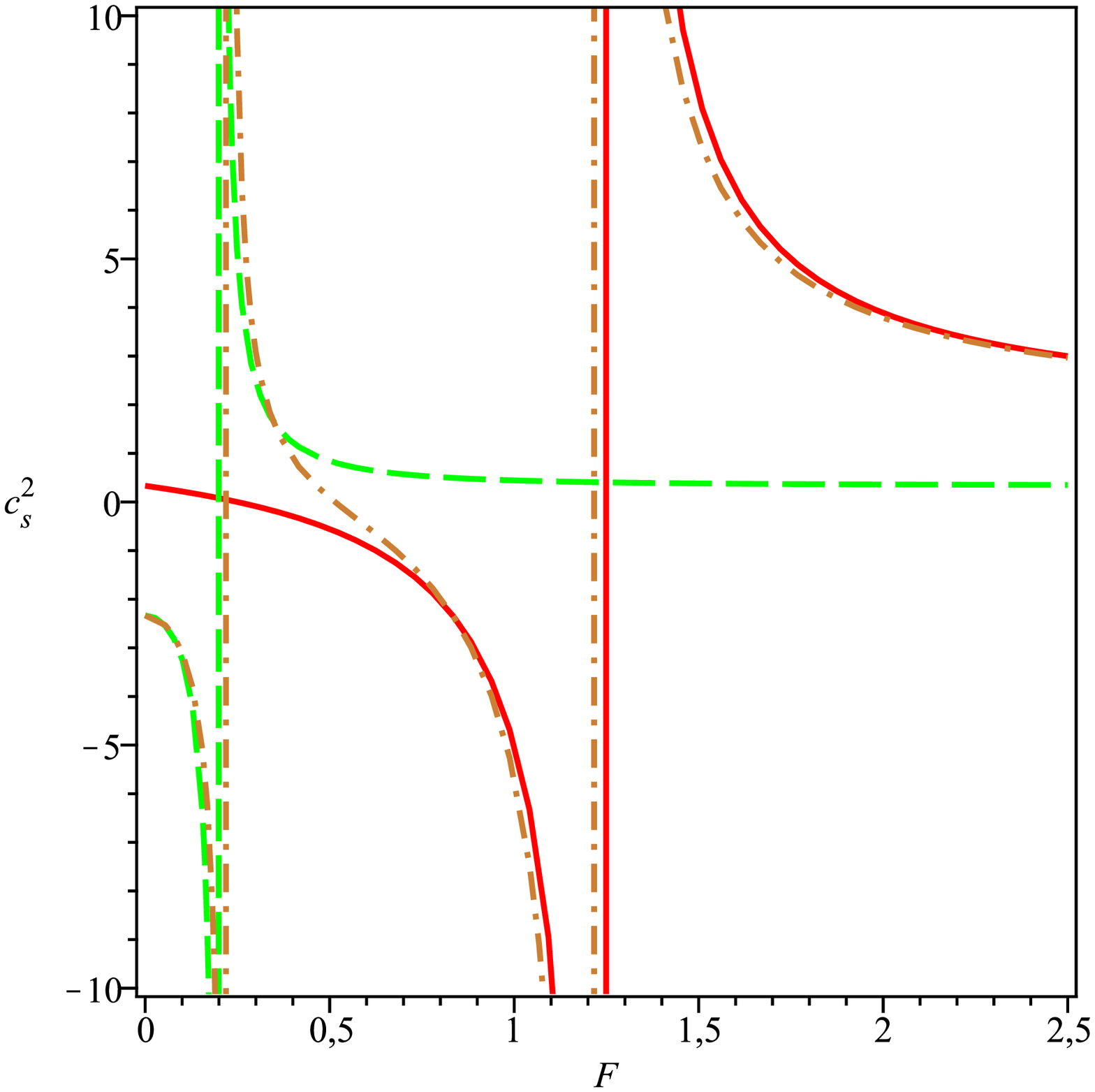}
\end{center}\vspace{0.3in}
\caption{Plot of the energy density of the magnetic field $\rho_\textsc{b}$ vs $F$ -- top panels -- and of the square sound speed $c_s^2(F)$ -- bottom panels -- for arbitrarily chosen $(\alpha,\gamma)$ [$(0.01,0.1)$ -- left-hand panels, $(0.1,0.1)$ -- center panels, and $(0.1,0.01)$ -- right-hand panels] for the different NLED models. The solid curve is for the model given by the Lagrangian density (\ref{l-f2}), while the dashed curve is for the model (\ref{l-1/f}), and the dash-dot is for the hybrid model which is based in the Lagrangian density ${\cal L}$ of Eq. (\ref{l-nled}).}\label{fig4}\end{figure*}


\section{Hybrid cosmic history}\label{sec-nled}

The authors of the reference \cite{novello-cqg-2007} proposed an hybrid model which interpolates between (\ref{l-f2}) and (\ref{l-1/f}). It was given by the Lagrangian density (\ref{l-nled}), which is a combination of the above Lagrangians. Here the quadratic term $\propto F^2$ dominates in very early epochs, the Maxwell term $\propto-F$ dominates in the radiation era, while the last term $\propto-F^{-1}$ is responsible for the accelerated phase. As for the model (\ref{l-1/f}), this model does not have the standard linear weak field Maxwell limit \cite{novello-cqg-2007}. 

The associated energy density of the magnetic field

\bea \rho_\textsc{b}=-\frac{4\alpha F^3-F^2-4\gamma}{4F},\label{rho-nled}\eea while the corresponding parametric pressure

\bea p_\textsc{b}=-\frac{20\alpha F^3-F^2+28\gamma}{12 F}.\label{p-nled}\eea The energy density has extrema at $F$-s which are roots of the algebraic equation $$\frac{d\rho_\textsc{b}}{dF}=-\frac{8\alpha F^3-F^2+4\gamma}{4F^2}=0;$$ 

\bea &&F_0=\frac{k^2+k+1}{24\alpha k},\nonumber\\
&&F_\pm=\frac{-(k-1)^2\pm\sqrt{3}i(k^2-1)}{48\alpha k},\label{roots}\eea where $$k=k(\alpha,\gamma):=\sqrt[3]{1-3456\gamma\alpha^2+48\alpha\sqrt{3\gamma(1728\gamma\alpha^2-1)}}.$$

Depending on the values of the free parameters $\alpha$, $\gamma$ there can be two extrema of $\rho_\textsc{b}$ -- a maximum and a minimum -- or none. The sign of $$\frac{d^2\rho_\textsc{b}}{dF^2}=-\frac{2\alpha F^3-2\gamma}{F^3},$$ determines whether the given extremum -- provided it exists -- is a maximum or a minimum. This is clearly seen from the top panels of FIG. \ref{fig4}, where the energy density of the magnetic field $\rho_\textsc{b}$ is plotted against $F=2B^2$, for different choices of the free parameters $(\alpha,\gamma)$: $(0.01,0.1)$ - LHP, $(0.1,0.1)$ - center panel, and $(0.1,0.01)$ - RHP. In the figure the model (\ref{l-f2}) is represented by the solid curve, while the model (\ref{l-1/f}) is depicted by the dashed curve, and the dash-dotted curve is for the hybrid model (\ref{l-nled}). In all cases the hybrid model interpolates between (\ref{l-f2}) at large values of the EM invariant $F$, and (\ref{l-1/f}) at small $F$-s. 

Since at large $F$-s the term $\propto 1/F$ may be neglected, then in the model (\ref{l-nled}) there is also an upper bound on $F$ which, depending on the free parameters [$\alpha$, $\gamma$], may or may not coincide with the one for the model (\ref{l-f2}) [$F=1/4\alpha$] but, in any case, is very close to it. In the present case, as in (\ref{l-f2}), there is a bounce at the upper bound of $F$, which is correlated with a local maximum of the background energy density $\rho_\textsc{b}$. Basically, the behavior in the neighborhood of the bounce -- both, prior to and after the bounce -- is very much like the one explained in section \ref{sec-f2}. However, in the distant past before the bounce -- during the contracting phase -- and in the future after it, the pace of the cosmic evolution is dictated by the term $\propto F^{-1}$ (for details see the former section \ref{sec-1/f}).\footnote{There are quite short stages before the bounce and after it, where the Maxwellian term drives a standard radiation era.} Hence, although the hybrid model is free of the big bang singularity, the cosmic evolution starts a finite time in the past in a cosmological singularity which, for the contracting universe, is the mirror image of the big rip in reversed time. Then the size of the universe decreases until the bounce occurs, to start growing until -- a finite time into the future -- the big rip singularity described in section \ref{sec-1/f} develops, meaning the catastrophic end of the cosmic evolution \cite{novello-cqg-2007}.

Troublesome as they are, the above curvature singularities -- which bound the entire cosmic history into a finite interval of time in the hybrid model (\ref{l-nled}) -- are less problematic than the insurmountable (classical) instabilities against small perturbations of the background energy density $\rho_\textsc{b}$, which develop whenever the square sound speed $c_s^2$ becomes a negative quantity, and the violations of causality associated with superluminal propagation of these perturbations. For the square sound speed in this model one has

\bea c_s^2=\frac{40\alpha F^3-F^2-28\gamma}{3(8\alpha F^3-F^2+4\gamma)}.\label{cs2-nled}\eea 

In the bottom panels of FIG. \ref{fig4}, $c_s^2$ in Eq. (\ref{cs2-nled}) is plotted against $F$ for the same values of the free parameters as before. As seen, at small $F$-s, $c_s^2$ in the hybrid model -- dash-dotted curve -- is always a negative quantity. Besides, right before the bounce (and after it), where in the model (\ref{l-1/f}) $c_s^2$ -- dashed curve -- has a vertical asymptote, the square sound speed (\ref{cs2-nled}) becomes a very large negative quantity. In all cases there are intervals, both at small and at large $F$-s, where $c_s^2>1$, implying obvious violations of causality in the model.


\begin{figure*}\begin{center}
\includegraphics[width=7cm,height=6cm]{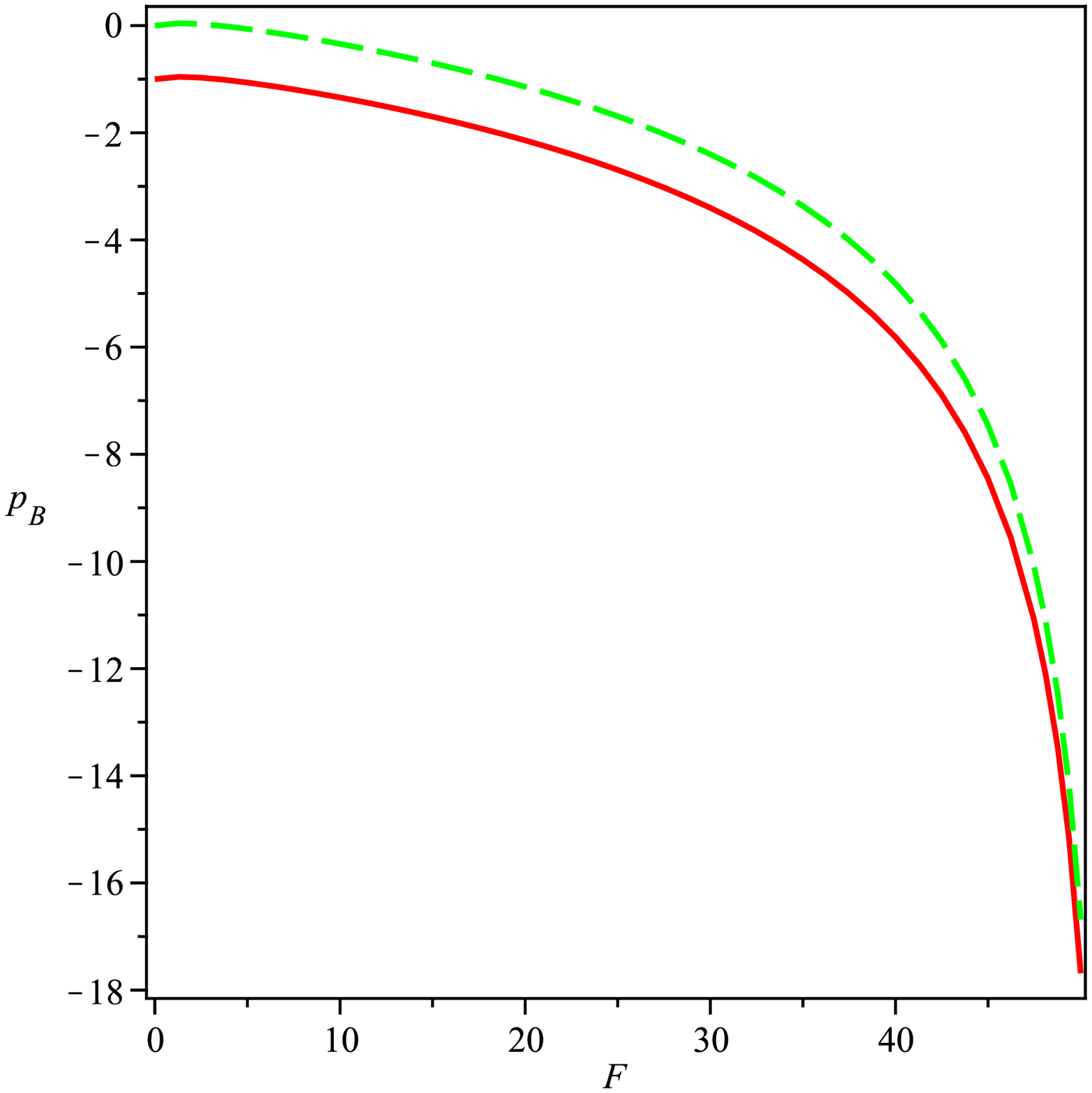}
\includegraphics[width=7cm,height=6cm]{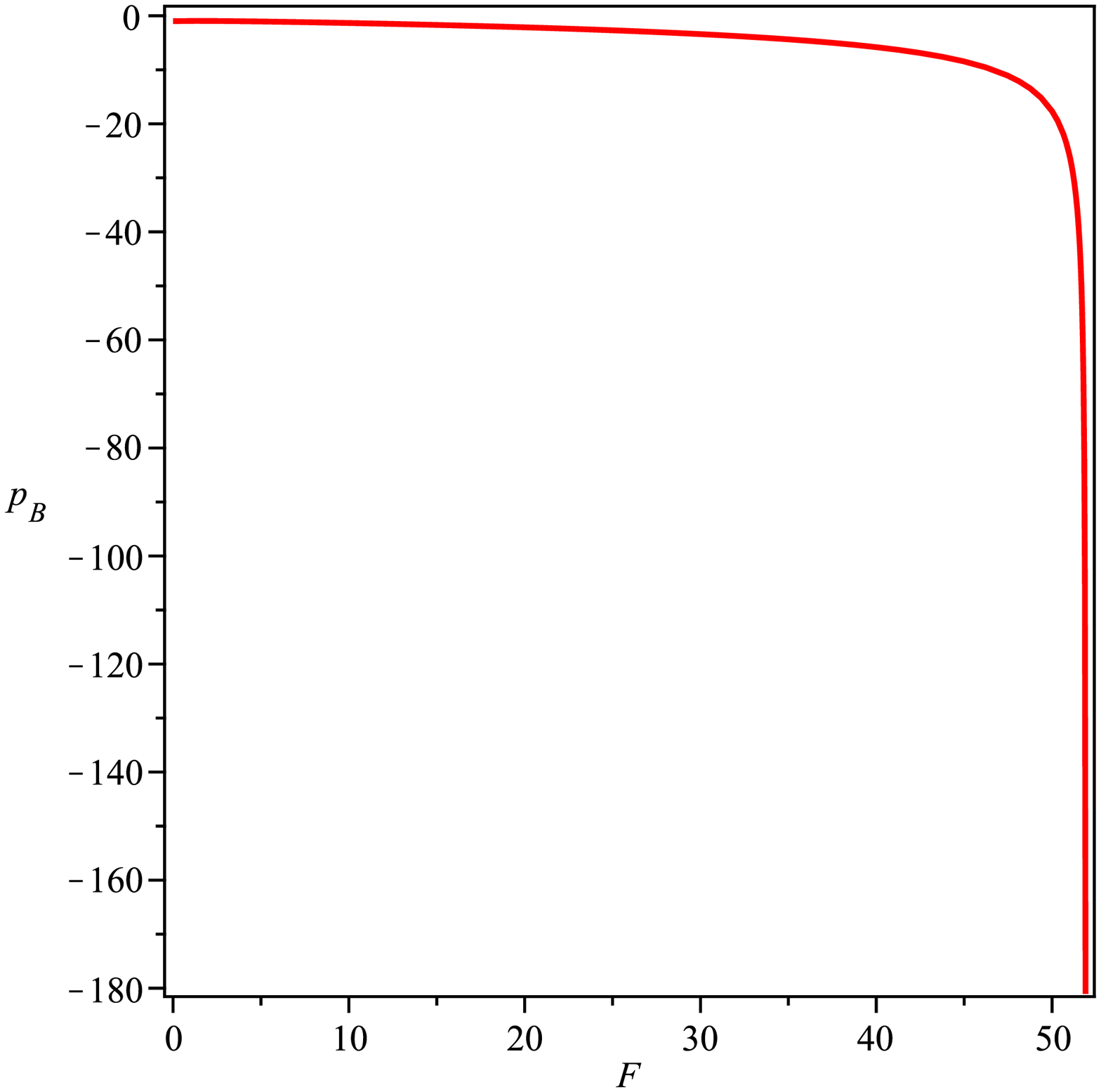}
\end{center}\vspace{0.3in}
\caption{Plots of the parametric pressure of the magnetic field $p_\textsc{b}$ vs $F$, for the models (\ref{l-model}) -- solid curve, and (\ref{l-model'}) -- dashed curve, for chosen values of the free parameters $(\alpha^2,\gamma^2)=(0.01,1)$. In the right-hand panel only $p_\textsc{b}=p_\textsc{b}(F)$ for the model (\ref{l-model}) is plotted. For the model (\ref{l-model'}), since $0\leq F\leq 1/2\alpha^2\gamma^2=50$, at the upper bound of the $F$-invariant ($F=50$), the pressure is a finite (negative) quantity. As shown in the right-hand panel, for the model (\ref{l-model}), since $0\leq F\leq F_+=51.9258$, as one approaches to $F_+$ the absolute value of the (negative) pressure $|p_\textsc{b}|$ unboundedly grows up.}\label{fig6}\end{figure*}


\section{Born-Infeld theory and its modifications}\label{sec-b-i}

Here we shall show that the BI Lagrangian (\ref{l-b-i}) and its modifications (\ref{l-model}) and (\ref{l-model'}), are also problematic and that no compelling cosmological model can be obtained out of them. We start with the Lagrangian (\ref{l-b-i}). This was studied in \cite{born-infeld} with emphasis in a static electric field with spherical symmetry -- corresponding to a charged body -- generating a regular electric field configuration without singular behavior. The gravitational contribution was not considered by the authors. However, as long as gravity is involved, as it is the case in cosmological settings, and magnetic universes are considered, the Lagrangian (\ref{l-b-i}) itself is not of interest since -- as long as $F\propto a^{-4}$, and given that $F$ is unbounded in the theory (\ref{l-b-i}) -- the big bang singularity develops in the corresponding cosmological model. An appropriate -- and subtle -- modification of (\ref{l-b-i}) can be given by the following Lagrangian density:

\bea {\cal L}=-\gamma^2\left(1-\sqrt{1-F/2\gamma^2}\right).\label{l-b-i-mod}\eea This theory has the correct linear weak field Maxwell limit and the field invariant $F$ is a bounded quantity: $0\leq F\leq 2\gamma^2$. This amounts to removing the big bang singularity from the corresponding cosmological model.

The energy density of the magnetic field $\rho_\textsc{b}$ and the corresponding parametric pressure $p_\textsc{b}$ are given by

\bea &&\rho_\textsc{b}=\gamma^2\left(1-\sqrt{1-F/2\gamma^2}\right),\nonumber\\
&&p_\textsc{b}=\gamma^2\left(\frac{1-F/6\gamma^2}{\sqrt{1-F/2\gamma^2}}-1\right).\label{rho-p-b-i}\eea While $\rho_\textsc{b}$ in (\ref{rho-p-b-i}) has no extrema, in the interval $0\leq F\leq 2\gamma^2$ it is a bounded quantity $0\leq\rho_\textsc{b}\leq\gamma^2$. The fact that $\rho_\textsc{b}$ is always a finite quantity -- as well as $H^2$ -- does not mean that the cosmological model based in (\ref{l-b-i-mod}) is free of curvature singularities. Actually, at the upper bound $F_*=2\gamma^2$ -- where the energy density of the magnetic field attains also an upper bound $\rho^*_\textsc{b}=\gamma^2$ -- the parametric pressure $p_\textsc{b}$ blows up. Besides, at $F_*$ the scale factor is a finite quantity $a(t)=a_*=(F_0/2\gamma^2)^{1/4}$. This means that at $F_*$ a sudden curvature singularity develops.

The square sound speed for the model (\ref{l-b-i-mod}) is $$c_s^2=\frac{1}{3}\left[\frac{2\gamma^2+F}{2\gamma^2-F}\right].$$ As long as $F\leq 2\gamma^2$ in this model, $c_s^2$ is always a positive quantity so that the stability issue does not arise. However, at $F_*$ the square sound speed is a vertical asymptote, i. e., as $F\rightarrow 2\gamma^2$, $c_s^2\rightarrow\infty$, it grows up without bounds. This fact may raise serious causality issues in the $F$-domain where the nonlinear effects are supposed to be dominating in this theory. 

A modification of the BI Lagrangian (\ref{l-b-i}) was proposed in the reference \cite{novello-prd-2012}. With the hope to adequate the BI theory to the case of a magnetic universe with the field invariant $F$ bounded , the authors added a term quadratic in $F$ within the square root in (\ref{l-b-i}) and, besides, removed the term $\propto\gamma^2$, resulting in the following Lagrangian density: $${\cal L}=-\gamma^2\sqrt{1+F/2\gamma^2-\alpha^2 F^2}.$$ 

As clearly seen this theory does not contain the classical linear Maxwell limit, since at weak field $${\cal L}\approx -\gamma^2-F/4.$$ Even if the dynamics of the classical EM fields is not modified by the term $\propto-\gamma^2$, quantum aspects related with the zero-point (vacuum) fluctuations of these fields are indeed included, which amount to a cosmological constant in a cosmological context. Since $W$ in Eq. (\ref{l-model}) is to be real, then the EM invariant $F$ is constrained to take values in the finite interval $0\leq F\leq F_+$, where $$F_+=\frac{1+\sqrt{1+16\alpha^2\gamma^4}}{4\alpha^2\gamma^2}.$$

The energy density of the magnetic field in this model $\rho_\textsc{b}=-{\cal L}$, vanishes at $F=F_+$ hence, the field invariant $F$ takes values in the finite interval $0\leq F\leq F_+$. Since $F=F_0/a^4$, the above means that the scale factor of the universe never vanishes, i. e., there is no big bang singularity in this model. The energy density of the magnetic field is a maximum at $$F=F_c=\frac{1}{4\alpha^2\gamma^2}\;\Rightarrow\;\rho_\textsc{b}^\text{max}=\frac{\sqrt{1+16\alpha^2\gamma^4}}{4\alpha},$$ while at vanishing $F=0$, it is non vanishing [$\rho_\textsc{b}(0)=\gamma^2$], due to inclusion of the energy density of vacuum field fluctuations through the term $-\gamma^2$ in the Lagrangian density. The parametric pressure $$p_\textsc{b}=-\frac{\gamma^2}{3}\frac{(3+F/2\gamma^2+\alpha^2 F^2)}{W^{1/2}},$$ where $W=1+F/2\gamma^2-\alpha^2 F^2$ was defined in (\ref{l-model}), is always negative. Notice that, at vanishing $F=0$, $$p_\textsc{b}=-\gamma^2=-\rho_\textsc{b},$$ i. e., the magnetic fluid behaves as a cosmological constant. This was expected in this model since, as explained above, at the linear weak field limit the present theory behaves as Maxwell theory plus a cosmological constant. At the upper bound $F=F_+$, where the nonlinear effects dominate the dynamics of this model, since $W=0$, then the pressure $p_\textsc{b}$ blows up -- see the FIG. \ref{fig6} -- while the scale factor of the universe at this field value $$F_+=\frac{F_0}{a^4}\;\Rightarrow\;a=\left(\frac{4\alpha^2\gamma^2 F_0}{1+\sqrt{1+16\alpha^2\gamma^4}}\right)^{1/4},$$ is a finite quantity. This means that a sudden curvature singularity -- not better than the big bang -- develops due to the nonlinear effects in this theory. But this is not the worst news for this model.


\begin{figure*}\begin{center}
\includegraphics[width=7cm,height=6cm]{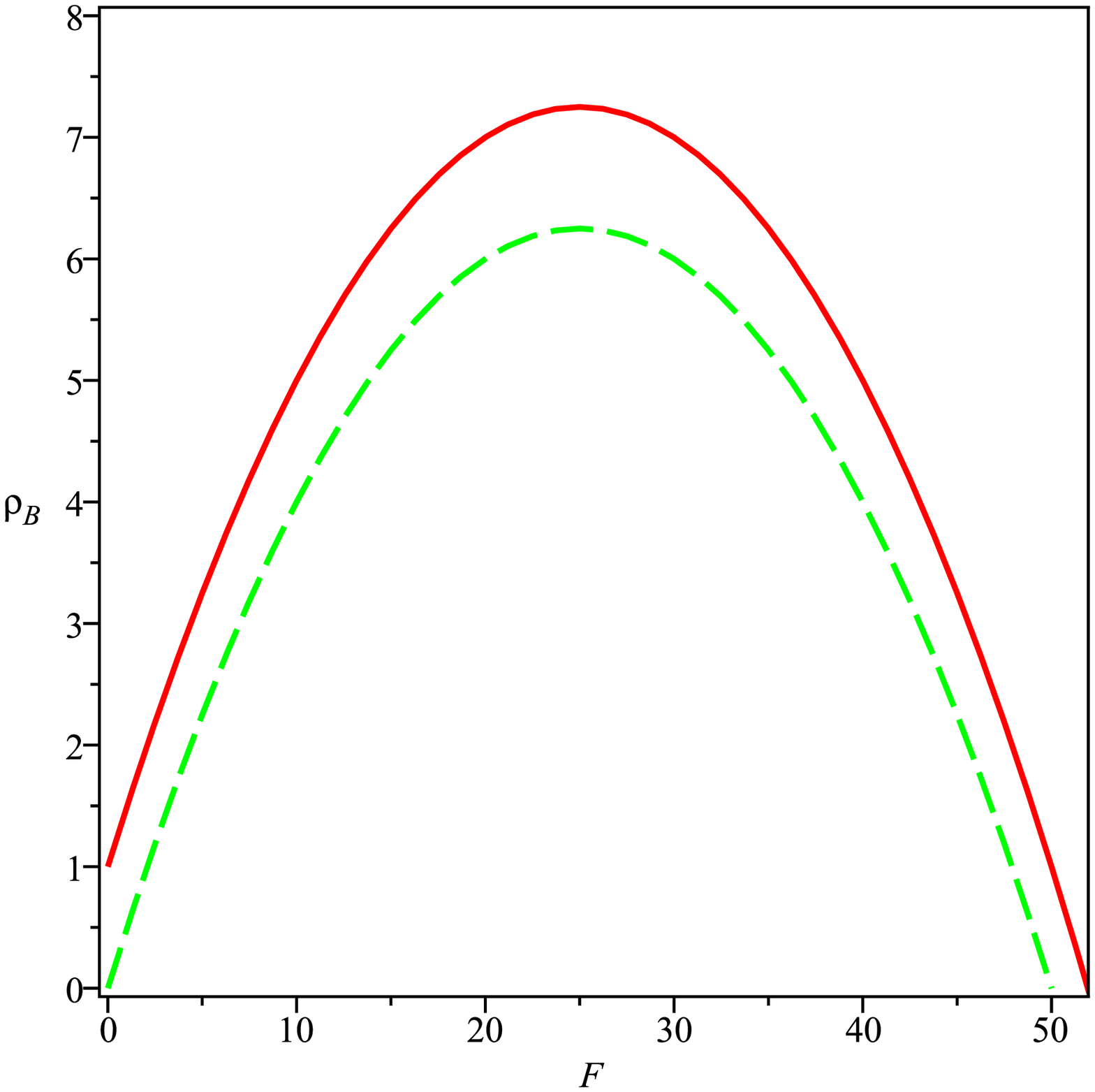}
\includegraphics[width=7cm,height=6cm]{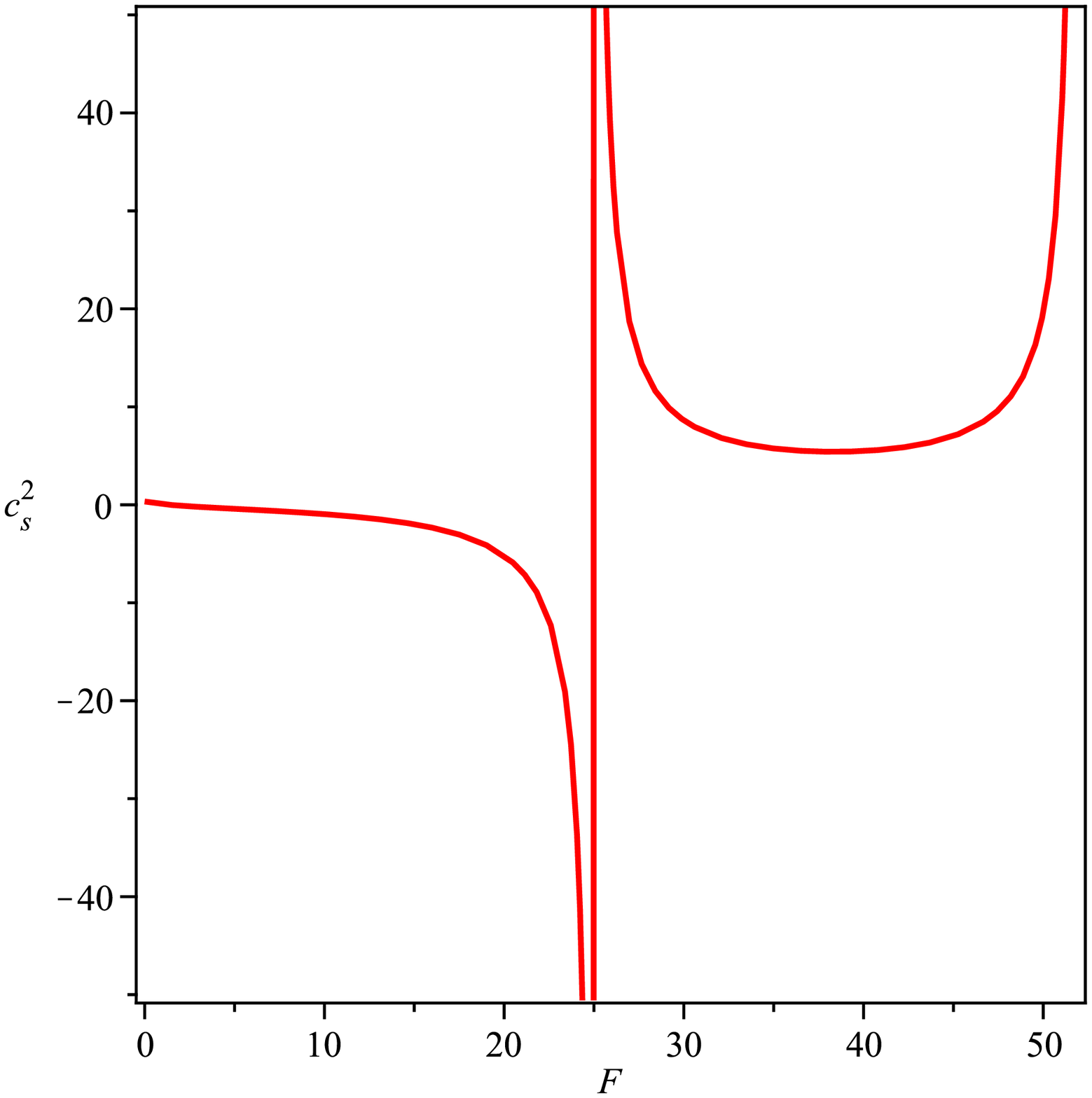}
\end{center}\vspace{0.3in}
\caption{Plots of the energy density of the magnetic field $\rho_\textsc{b}$ -- left-hand panel -- for the models (\ref{l-model}) -- solid curve, and (\ref{l-model'}) -- dashed curve, and of the square sound speed $c^2_s$ (exactly the same for both models) -- right-hand panel -- for arbitrarily chosen values of the free parameters $(\alpha^2,\gamma^2)=(0.01,1)$. It is seen that, at $F=F_c$, where the energy density of the magnetic field is a maximum, the square sound speed has a vertical asymptote. To the left of the asymptote there is an interval $\Delta F$, where $c_s^2$ is a negative quantity. Besides, as one approaches to $F_c$ from the left, $c_s^2\rightarrow-\infty$. Between the asymptotes $F_c<F<F_+$, the square speed of sound $c_s^2>1$.}\label{fig5}\end{figure*}


The square sound speed is given by 

\bea c_s^2=\frac{1}{3\gamma^2}\left[\gamma^2-\frac{(1+16\alpha^2\gamma^4)F}{W(1-4\alpha^2\gamma^2 F)}\right].\label{s-speed}\eea 

It is seen from this expression that at vanishing field $F=0$, $c_s^2=1/3$. In general, while for $0\leq F\leq F_r$, where $F_r$ is the positive real root of the cubic equation $$8\alpha^4\gamma^4F^3-6\alpha^2\gamma^2F^2-(1+40\alpha^2\gamma^4)F+2\gamma^2=0,$$ the square sound speed meets the required bounds $1>1/3\geq c_s^2\geq 0$, for $F_r<F<F_c=1/4\alpha^2\gamma^2$ it is a negative quantity. Another important feature seen from Eq. (\ref{s-speed}) is that the square sound speed has vertical asymptotes: (i) at $F=F_c$, where the energy density of the magnetic field is a maximum, and (ii) at the upper bound $F=F_+$. This is illustrated in the RHP of the FIG. \ref{fig5}, where a plot of $c_s^2$ vs $F$ is shown for arbitrarily chosen values of the free parameters. As one approaches to the asymptote at $F=F_c$ from the left -- while $F$ takes values in the interval $F_r<F<F_c$ -- the square sound speed takes increasingly negative values, $$\lim_{F\rightarrow F^-_c} c_s^2=-\infty,$$ thus causing insurmountable classical instability of small perturbations of the background. In the interval between the asymptotes ($F_c<F<F_+$) the square sound speed is positive and always exceeds the local speed of light squared ($c_s^2>1$).\footnote{It is not difficult to show that, for $F$-s in the interval $F_c<F<F_+$, $c_s^2$ is a minimum at the real positive root $F=F_\text{min}$ of the cubic equation $8\alpha^2\gamma^2F^3-3F^2-1/\alpha^2=0$: $c_{s,\text{min}}^2=c_s^2(F_\text{min})$. It can be shown that $$\lim_{\alpha\gamma^2\rightarrow 0}c^2_{s,\text{min}}=17/3>1,\;\lim_{\alpha\gamma^2\rightarrow\infty}c^2_{s,\text{min}}=5/3>1.$$ This means that, for $F_c<F<F_+$, independent of the values of the free parameters $\alpha$, and $\gamma$, the minimum value of the square sound speed always exceeds the speed of light.} This may raise serious causality issues \cite{roy}. 

A modification of the BI Lagrangian (\ref{l-b-i}) which keeps the spirit of (\ref{l-model}) but which -- unlike this latter Lagrangian -- does actually has the classical linear weak field Maxwell limit, is the Lagrangian density (\ref{l-model'}). The energy density of the magnetic field in this case is given by 

\bea \rho_\textsc{b}=\gamma^2\left(W^{1/2}-1\right).\label{rho-b-model'}\eea It vanishes at vanishing field [$F=0$] as it should be for an EM theory which respects the linear Maxwell limit. The energy density of the magnetic field in Eq. (\ref{rho-b-model'}) is non negative in the interval $$0\leq F\leq\frac{1}{2\alpha^2\gamma^2}<F_+.$$ Hence the field $F$ is bounded to this interval in this model. The scale factor of the universe does never vanish, meaning that the model is free of the big bang singularity. 

The parametric pressure of the magnetic field is depicted by the following equation: $$p_\textsc{b}=-\gamma^2\left(\sqrt{W}-1\right)+\frac{(1-4\alpha^2\gamma^2 F)F}{3\sqrt{W}}.$$ It is always a finite quantity since the upper bound on the values $F$ can take in this model [$F=1/2\alpha^2\gamma^2$] is always below the value $F_+$, at which $W$ vanishes. Hence, this model is also free of curvature singularities of the sudden type. Nevertheless, it is not free of the serious stability and causality problems originated from violations of the bounds $0\leq c_s^2\leq 1$. Actually, in this model $c_s^2$ exactly coincides with the one for the model (\ref{l-model}) which is given by Eq. (\ref{s-speed}). These only differ in that the vertical asymptote at $F=F_+$ -- which is present in the model (\ref{l-model}) -- in this case does not arise. Notwithstanding, the asymptote at $F_c=1/4\alpha^2\gamma^2$ is still there, and the problem with negativity of the square sound speed -- for $F$-s to the left of the asymptote -- still stands, as well as the causality problem associated with $c_s^2>1$ to the right of the asymptote. Just like in the case (\ref{l-model}).

\section{Discussion and Conclusions}\label{sec-disc}

Although experiment (observations) is the one who ultimately decides whether a given physical theory is right or wrong, there are a few physical principles on which the fundamental theories of physics are grounded, which -- conventional wisdom states -- should be always satisfied. Among them causality plays a special role. Besides, physical theories which are intended to describe our present universe, should be stable against small perturbations of the background since, otherwise, the world as we see it would be the result of pure chance and not of the joint synchronized action of the fundamental laws of physics. One may think of these basic principles as a kind of coarse filter for plausible theories, while experimental/observational testing represents the finest possible such filter. In a cosmological context one has, for instance, the type Ia supernovae and $H(z)$ data tests \cite{data-test}. Any feasible cosmological model has to pass these tests (among others). But, what if prior to testing a given cosmological model by means of the sophisticated techniques which are involved in the data analysis (see, for instance, \cite{jcap-2013, nora-jcap}), one performs a simple check of the mentioned basic principles? The result might be unexpected.

In the present paper we have performed a simple check of stability and causality of several NLED-based theories, and we have shown that none of them can be a plausible cosmological model. Since, under the assumptions made here, the magnetic universe can be pictured as an homogeneous and isotropic FRW spacetime filled with a purely magnetic fluid, small perturbations of the background should propagate at subluminal (at most luminal) local speed if one wants to avoid violations of causality. Besides, since one expects these cosmological models to be classically stable -- otherwise they would not be feasible cosmological models capable of describing a long lasting stage of the cosmic evolution -- then, the small fluctuations of the background energy density should obey the wave equation $\delta\ddot\rho=c_s^2\nabla^2\delta\rho$, with $c_s^2\geq 0$. The following bounds on the square sound speed are to be satisfied: $0\leq c_s^2\leq 1$. It is surprising that such a simple and basic check can serve to reject several cosmological models as unphysical, which, otherwise, seem to be adequate models of our cosmos.

Take, for instance, the Born-Infeld Lagrangian (\ref{l-b-i}) \cite{born-infeld}. According to (\ref{speed-sound}), the (square) speed at which small fluctuations in a magnetic universe propagate is given by $$c_s^2=\frac{2\gamma^2-F}{3(2\gamma^2+F)}.$$ For $F>2\gamma^2$, i. e., at large field values (recall that in this case the EM invariant $F$ is unbounded from above), $c_s^2<0$, which means that the model is classically unstable. For other modifications of BI-theory (\ref{l-b-i-mod}), (\ref{l-model}), and (\ref{l-model'}) -- based on the simple analysis of the square sound speed -- not only classical instability, but also obvious violations of causality are present. A similar story takes place for other less sophisticated Lagrangian densities (\ref{l-f2}), (\ref{l-1/f}), and (\ref{l-nled}). None of the above mentioned cosmological models (\ref{l-b-i}-\ref{l-nled}) meets the required bounds $0\leq c_s^2\leq 1$. Hence, these are to be rejected as unphysical. Perhaps the addition of other cosmological matter fields might improve this situation. We leave this for further investigation.

Our conclusion is that, before pushing any further a given cosmological model (experimental/observational testing, study of physical implications, etc.), one should perform a simple check of the basic principles of physics. This can save time and effort. 

It could be interesting to discuss the stability and causality criteria illustrated in this paper also in the context of other bouncing models which are not based on NLED as, for instance, in the context of string cosmology models where the bounce is due to a non-local dilaton potential \cite{gasperini}.

The authors thank Nora Breton for useful comments and SNI-CONACYT Mexico for support. The work of R G-S was partly supported by SIP20131811, COFAA-IPN, and EDI-IPN grants.

\end{document}